%% file: GormleyFruhwirthSchnatter.tex
\documentclass[a4paper,12pt]{article}
\usepackage{authblk}
\usepackage[colorlinks,citecolor=blue,linkcolor=blue,urlcolor=blue,filecolor=blue]{hyperref}
\usepackage[margin=0.9in]{geometry}

\usepackage{tikz}
\usepackage{textcomp}
\usepackage{fixltx2e,fix-cm}
\usepackage{amssymb}
\usepackage{amsmath}
\usepackage{enumerate}
\usepackage{graphics}
\usepackage{graphicx}
\usepackage{subfigure}
\usepackage{makeidx}
\usepackage{multicol}
\usepackage{natbib}
\usepackage{xspace}
\usepackage{nicefrac}
\usepackage{algorithm}
\usepackage{afterpage}
\usepackage{algorithmic}
\usepackage{relsize}
\usepackage{threeparttable}
\usepackage{bm,bbm,xr}
\usepackage{xr}
\usepackage{xspace}
\usepackage{bibunits}
\usepackage{xcolor}

\input tikzlibrarybayesnetcode.tex

\usepackage[T1]{fontenc}
\usepackage{lmodern}

\newcommand{\thmod}{\theta}
\newcommand{\ym}{y}
\newcommand{\bfz}{0}
\newcommand{\identm}{I}
\newcommand{\muswt}[2]{\mu_{#1,#2}}
\newcommand{\Xbeta}{x}
\newcommand{\Xbetamat}{X}
\newcommand{\ndes}{p}
\newcommand{\sigmaerrsw}[1]{\sigma^2_{ #1}}
\newcommand{\sigmaerrsqrtsw}[1]{\sigma _{ #1}}
\newcommand{\betacsw}[2]{\beta_{#1,#2}}
\newcommand{\betavsw}[1]{\beta_{#1}}
\newcommand{\logit}{\mbox{\rm logit }}
\newcommand{\xiv}{\xi}
\newcommand{\Lold}{q}
\newcommand{\dold}{D}
\newcommand{\Xbetatilde}{\tilde{\Xbeta}}
\newcommand{\idestar}[1]{#1 ^{\star}} 
\newcommand{\idestararg}[2]{#1 ^{#2,\star}} 
\newcommand{\BinoT}{N}  
\newcommand{\pl}{\pi}  
\newcommand{\rvY}{Y} 
\newcommand{\plt}[1]{\pl_{#1}} 
\newcommand{\ydens}{y}  
\newcommand{\Bincoef}[2]{\left( \begin{array}{c}   #1 \\#2  \end{array}\right)}
\newcommand{\fnarg}[2]{h_{#1} (#2)}  
\newcommand{\true}{^{\rm true}}
\newcommand{\samspace}{\mathcal{Y}} 
\newcommand{\ideU}{U}
\newcommand{\Prob}{\mbox{\rm P}}
\newcommand\automaticrules{\@tablerulestrue}
\newcommand\noautomaticrules{\@tablerulesfalse}
\newcommand{\zm}{\mathbf{z}} 
\newcommand{\lik}{{\text{L}}\xspace} 
\newcommand{\likc}{\lik_\text{c}} 
\newcommand{\indic}[1]{\mathbb{I} (#1)}
\newcommand{\Normal}{\mathcal{N}}
\newcommand{\Wishartinv}{\mathcal{IW}}
\newcommand{\Mulnom}{\mathcal{M}}
\newcommand{\um}{{\mathbf u}}
\newcommand{\rmm}{{\mathbf r}}
\newcommand{\perm}{\sigma}
\newcommand{\im}[1]{^{(#1)}} 
\newcommand{\hist}{{\cal H}}
\newcommand{\Dir}{\mathcal{D}}
\newcommand{\Ew}{\mbox{\rm E}}
\newcommand{\wm}{{\mathbf w}}
\newcommand{\Bino}{\mathcal{B}}
\newcommand{\Uniform}{\mathcal{U}}
\newcommand{\IG}{\mathcal{IG}}

\begin{document}
\title{\textbf{Mixtures of Experts Models}\thanks{A chapter prepared for the forthcoming \emph{Handbook of Mixture Analysis}}}

\author{Isobel Claire Gormley\thanks{School of Mathematics and Statistics, Insight Centre for Data Analytics, University College Dublin, Ireland. \url{claire.gormley@ucd.ie}}}
\author{Sylvia Fr\"uhwirth-Schnatter \thanks{Institute for Statistics and Mathematics, Vienna University of Economics and Business, Austria. \url{sfruehwi@wu.ac.at}}}
\affil[]{}

\date{}
\maketitle
\begin{abstract}
	Mixtures of experts models provide a framework in which covariates may be included in mixture models. This is achieved by modelling the parameters of the mixture model as functions of the concomitant covariates. Given their mixture model foundation, mixtures of experts models possess a diverse range of analytic uses, from clustering observations to capturing parameter heterogeneity in cross-sectional data. This chapter focuses on delineating the mixture of experts modelling framework and demonstrates the utility and flexibility of mixtures of experts models as an analytic tool.
\end{abstract}

\section{Introduction}
\label{sec:introCGSF}

The terminology \emph{mixtures of experts models} encapsulates a broad class of mixture models in which the model parameters are modelled as functions of concomitant covariates. While the response variable $y$ is modelled via a mixture model, model parameters are modelled as functions of other, related, covariates $x$ from the context under study.

The mixture of experts nomenclature (ME) has its origins in the machine-learning literature \citep{jacobs91}, but mixtures of experts models appear in many different guises, including switching regression models\index{switching regression models} \citep{quandt72}, concomitant variable latent-class models \citep{dayton1988}, latent class regression models\index{latent class regression} \citep{desarbo88}, and mixed models \citep{wang96}. \cite{li2011} discuss finite smooth mixtures, a special case of ME modelling. \cite{mclachlan:peel:2000} and \cite{fru:book} provide background to a range of mixtures of experts models; \cite{masoudnia2014} survey the ME literature from a machine learning perspective.

The mixture of experts framework facilitates flexible modelling, allowing a wide range of application. ME models for rank data \citep{gormley08b, gormley10}, ME models for network data \citep{gormley10b}, for time series data \citep{waterhouse1996, huerta2003, fruhwirth2012}, for non-normal data \citep{vil-etal:reg,chamroukhi2015} and for longitudinal data \citep{tang2015}, among others, have been developed. \cite{peng1996} employed a hierarchical mixture of experts model in a speech recognition context. The general ME framework has also been incorporated in the mixed membership model setting, giving rise to a mixed membership of experts model \citep{white2014}, and into the infinite mixture model setting \citep{rasmussen2002}.  Cluster weighted models \citep{ingrassia2015, subedi2013, gershenfeld1997} are also closely related to ME models.

This chapter introduces the generic mixture of experts framework, in Section~\ref{sec:MEframework}, and describes approaches to inference for ME models in Section~\ref{sec:inference}. A broad range of illustrative data analyses are given in Section~\ref{sec:apps}, and an overview of existing softwares which fit ME models is provided.   Section~\ref{sec:identifiability} discusses identifiability issues for   mixtures of experts models.
The chapter concludes with some discussion of the benefits and issues of the ME framework, and of some areas ripe for future development.

\section{The Mixture of Experts Framework}
\label{sec:MEframework}

Any mixture model which incorporates covariates or concomitant variables falls within the mixture of experts framework.

\subsection{A mixture of experts model}
\label{sec:MEdefn}

Let $y_1, \ldots, y_n$ be an independent and identically distributed sample of outcome variables from a population modelled by a $G$ component finite mixture model.  Depending on the application context, the outcome variable can be univariate or multivariate, discrete or continuous, or of a more general structure such as time series or  network data. Each component $g$ (for $g = 1, \ldots, G$) is modelled by the probability density function $f_g(\cdot | \theta_g)$ with parameters denoted  by $\theta_g$, and has weight $\eta_g$ where $\sum_{g=1}^{G} \eta_g = 1$. Observation $y_i$ ($i = 1, \ldots, n$) has $\Lold$ associated covariates, which are denoted $x_i$. The ME model extends the standard finite mixture model introduced in Chapter 1 of this volume by allowing model parameters to be functions of the concomitant variables $x_i$:
\begin{eqnarray}
p(y_i|x_i) & = & \sum_{g=1}^{G} \eta_g (x_i) f_g(y_i | \theta_g(x_i)).
\label{eqn:genericME}
\end{eqnarray}
ME models can be considered as a member of the class of conditional mixture models\index{conditional mixture models} \citep{Bishop06}; for a given set of covariates $x_i$, the distribution of $y_i$ is a finite mixture model. \cite{jacobs91} consider the component densities $f_g(y_i | \theta_g(x_i))$ as the \emph{experts}, which model different parts of the input space, and the component weights $\eta_g(x_i)$ as the \emph{gating networks},\index{gating networks} hence the \emph{mixture of experts} terminology.

\begin{figure}[t!]
\begin{center}	
\includegraphics[width=200pt, height=250pt, angle=270]{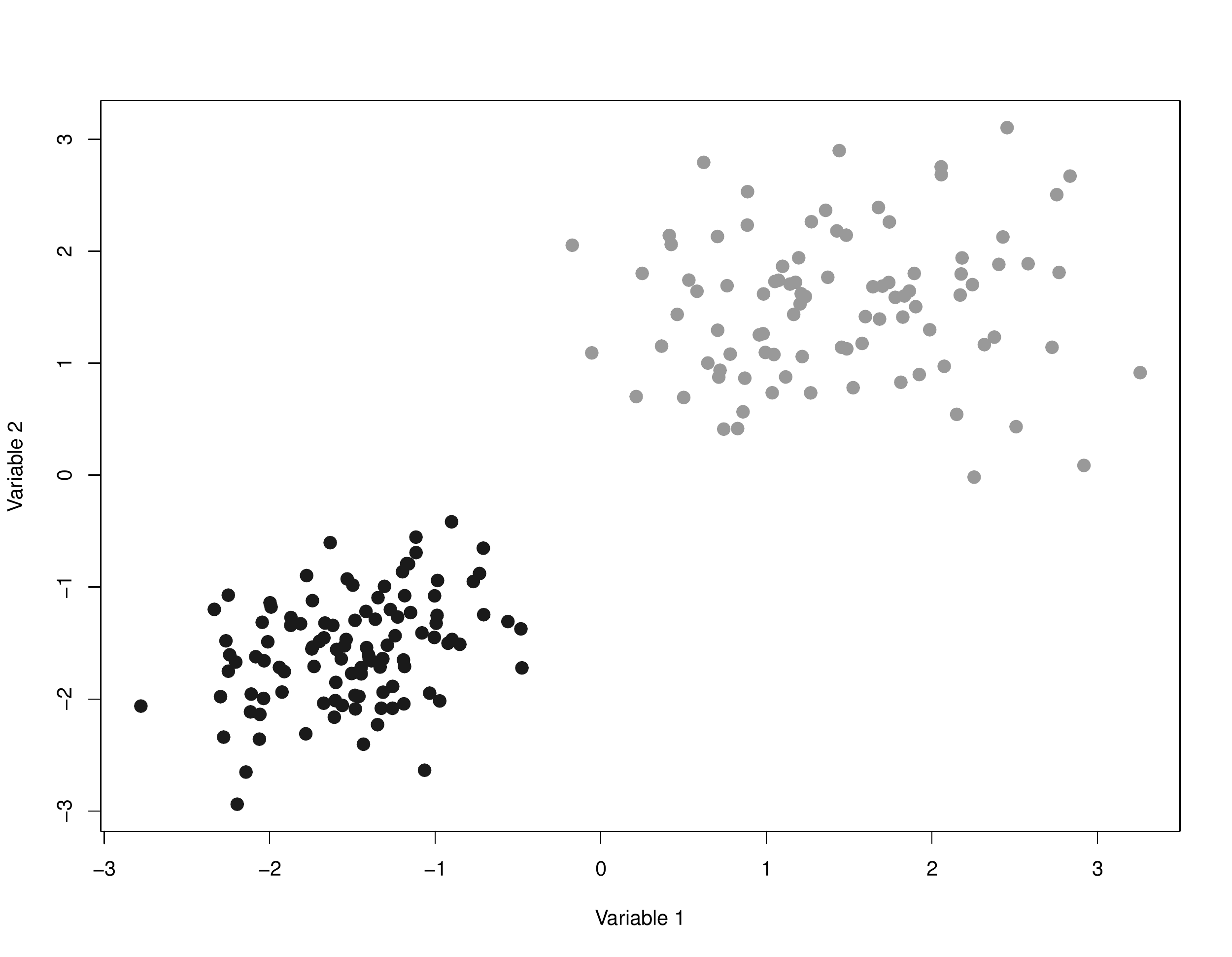}
\end{center}
\caption{A two-dimensional simulated data set, from a $G = 2$ ME model. Black observations belong to cluster 1, and grey to cluster 2, based on the MAP clustering from fitting a $G = 2$ mixture of bivariate Gaussian distributions.}
\label{fig:simdata}
\end{figure}

The models for $\eta_g(x_i)$ and for $\theta_g(x_i)$ in (\ref{eqn:genericME}) vary and are typically application specific. For example, \cite{jacobs91} model the component weights using a  multinomial logit  (MNL) regression model, and the component densities using  generalized linear models. \cite{young2010} provide further flexibility by allowing the mixing proportions to be modelled nonparametrically, as a function of the covariates.

\subsection{An illustration}
\label{sec:illustrativeeg}

A simple simulated data set is employed here to introduce the mixture of experts framework. Figure~\ref{fig:simdata} shows $n = 200$ two-dimensional continuously valued observations, $y_1, \ldots, y_n$, simulated from an ME model  with $G = 2$ components. A single ($\Lold = 1$) categorical covariate $x_i$ is associated with each observation representing, for example, gender where level 0 denotes female. Interest lies in clustering the observations and exploring any relations between the resulting clusters and the associated covariate.

It is common that a clustering method is implemented on the outcome variables of interest, $y_1, \ldots, y_n$, without reference to the covariate information. Once a clustering has been produced, the user typically probes the clusters to investigate their structure. Interpretations of the clusters are produced with reference to values of the model parameters within each cluster and with reference to the covariates that were not used in the construction of the clusters. Therefore, a natural approach to modelling the data in Figure~\ref{fig:simdata} is to cluster them by fitting a two component mixture of bivariate Gaussian distributions to $y_1, \ldots, y_n$. The \emph{maximum a posteriori} (MAP) cluster membership of each observation resulting from fitting such a model is also illustrated in Figure~\ref{fig:simdata}.

A cross tabulation of the MAP cluster memberships and the gender covariate is given in Table \ref{tab:gend}. It is clear that females have a strong presence in cluster 1, and males in cluster 2. However, the mixture of Gaussians model fitted does not incorporate or quantify this relationship or its associated uncertainty. It is in such a setting that an ME model is useful.

\begin{table}[t!]
\begin{center}
\caption{Cross tabulation of MAP cluster memberships and the gender covariate for the simulated data of Figure \ref{fig:simdata}}
\label{tab:gend}
\begin{tabular}{ccc}
 & Female & Male\\
 \hline
Cluster 1 & 75& 17\\
Cluster 2 & 23& 85\\
\end{tabular}
\end{center}
\end{table}

The model from which the data in Figure \ref{fig:simdata} are simulated is an ME model where $f_{g}(y_i|\theta_g) = 
 \phi(y_i| \mu_g, \Sigma_g)$ is the density of a bivariate Normal  distribution
and in which the component weights arise from a multinomial logit 
 model with $G$ categories with gender as covariate $x_i$,  i.e.
\begin{eqnarray}
\log\left[\frac{\eta_g(x_i)}{\eta_1(x_i)}\right] & = & \gamma_{g0} + \gamma_{g1} x_{i},
\label{eqn:logreg}
\end{eqnarray}
where cluster 1 is the baseline cluster  with $\gamma_1 = (\gamma_{10}, \gamma_{11})^\top = (0,0)^\top$, and $g=2,\ldots,G$. In our example, where $G=2$, model
(\ref{eqn:logreg}) reduces to a binary logit model. The parameter $\gamma_{g1}$ (and its associated uncertainty) quantifies the relationship between the gender covariate and  membership of cluster $g$,  with   $\gamma_{g1}=0$ corresponding to independence between cluster membership and the gender covariate. Note that such a model easily extends to $\Lold>1$ covariates $x_i = (x_{i1}, \ldots, x_{iq})$ with associated parameter $\gamma_g=(\gamma_{g0}, \ldots, \gamma_{g \Lold})^\top$ for cluster $g$.

Fitting such an ME model to the simulated data results in a MAP clustering unchanged from that reported in Table \ref{tab:gend} and gives the maximum likelihood estimate $\hat{\gamma}_{21} = 2.79$, with standard error $0.36$. (Details of the maximum likelihood estimation process and standard error derivation  follow in Section~\ref{sec:mle}.)  Thus, the odds of a male belonging to cluster 2 are $\exp(2.79) \approx 16$ times greater than the odds of a female belonging to cluster 2. Thus the ME model has clustering capabilities and provides insight into the type of observation which characterises each cluster.

\subsection{The suite of ME models}
\label{sec:allMEmodels}

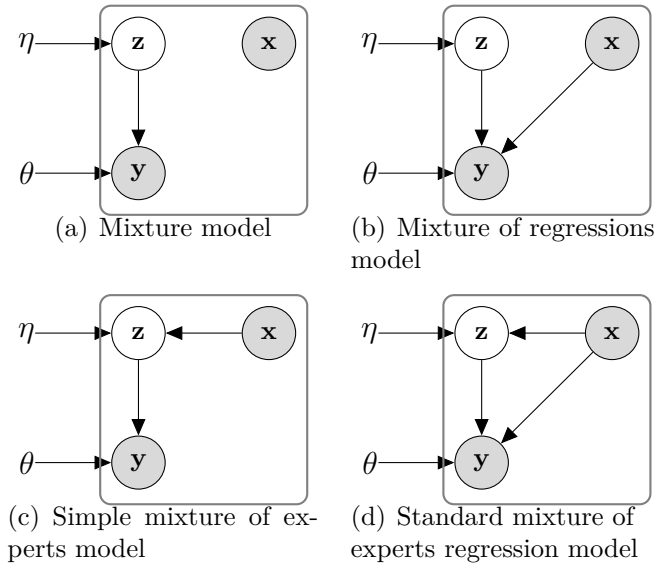
\begin{figure}[t!]
\begin{center}
\begin{tabular}{cc}
\subfigure[Mixture model]{\label{fig:MEa}
  \begin{tikzpicture}
  \node[obs, fill=gray!30] (y) {$\mathbf{y}$};
  \node[latent, above=of y, xshift=0cm] (z) {$\mathbf{z}$};
   \node[obs,fill=gray!30,right=of z](x){$\mathbf{x}$};
  \node[const, left=of y, xshift=0cm,yshift=0cm]  (theta) {$\theta$} ;
  \node[const, left=of z, xshift=0cm]  (eta) {$\eta$} ;
  \edge {z,theta} {y} ;
  \edge {eta} {z} ;
  \plate[color=gray, thick, minimum width=15ex] {py} {(y)(z)(x)} {} ;
\end{tikzpicture}
} &
\subfigure[Mixture of regressions \newline model]{\label{fig:MEb}
\begin{tikzpicture}
  \node[obs, fill=gray!30] (y) {$\mathbf{y}$};
  \node[latent,above=of y] (z) {$\mathbf{z}$};
  \node[const,left=of y]  (theta) {$\theta$} ;
  \node[const,left=of z]  (eta) {$\eta$} ;
 \node[obs,fill=gray!30,right=of z](x){$\mathbf{x}$};
  \edge {z,x} {y} ;
  \edge {eta} {z} ;
  \edge {theta}{y};
  \plate[color=gray, thick, minimum width=15ex] {py} {(y)(z)(x)} {} ;
\end{tikzpicture}
}\\

\subfigure[Simple mixture of experts model]{\label{fig:MEc}
  \begin{tikzpicture}
  \node[obs, fill=gray!30] (y) {$\mathbf{y}$};
 \node[obs,fill=gray!30,right=of z](x){$\mathbf{x}$};
  \node[latent, above=of y, xshift=0cm] (z) {$\mathbf{z}$};
  \node[const, left=of y, xshift=0cm,yshift=0cm]  (theta) {$\theta$} ;
  \node[const, left=of z, xshift=0cm]  (eta) {$\eta$} ;
  \edge {z,theta} {y} ;
  \edge {eta,x} {z} ;
  \plate[color=gray, thick, minimum width=15ex] {py} {(y)(z)(x)} {} ;
\end{tikzpicture}
}&
\subfigure[Standard mixture of \newline experts regression model]{\label{fig:MEd}
\begin{tikzpicture}
  \node[obs, fill=gray!30] (y) {$\mathbf{y}$};
  \node[latent, above=of y, xshift=0cm] (z) {$\mathbf{z}$};
  \node[const, left=of y, xshift=0cm,yshift=0cm]  (theta) {$\theta$} ;
  \node[const, left=of z, xshift=0cm]  (eta) {$\eta$} ;
   \node[obs,fill=gray!30,right=of z](x){$\mathbf{x}$};

  \edge {z,theta,x} {y} ;
  \edge {eta,x} {z} ;
  \plate[color=gray, thick, minimum width=15ex] {py} {(y)(z)(x)} {} ;
\end{tikzpicture}
}
\end{tabular}
\caption{The graphical model representation of mixtures of experts models. The differences between the four special cases are due to the presence or absence of edges between the covariates $x$ and the latent variable $z$ and response variable $y$. For model (a) $p(y,z|x)= p(y|z) p(z)$, for model (b) $p(y,z|x)= p(y|x,z) p(z)$, for model (c) $p(y,z|x)= p(y|z) p(z|x)$, whereas for model (d) $p(y,z|x)= p(y|x,z) p(z|x)$.}\label{fig:MEgraphical}
\end{center}
\end{figure}

The ME model outlined in Section~\ref{sec:illustrativeeg} involves modelling the component weights as a function of covariates.
This is one model type (termed a \emph{simple mixture of experts model}) from the ME framework. Figure \ref{fig:MEgraphical} shows a graphical model representation of the suite of four models in the ME framework,  based on a latent variable representation of the mixture model  (\ref{eqn:genericME}), involving the latent cluster membership of each  observation, denoted $z_i$,   where $z_i = g$ if observation $y_i$
belongs to cluster $g$. The indicator variable $z_i$ therefore has a multinomial distribution with a single trial and probabilities equal to $\eta_{g}(x_i)$ for $g = 1, \ldots, G$ and the latent variable representation reads:
\begin{eqnarray}
y_i |  x_i, z_i=g \sim   f_g(y_i | \theta_g(x_i)), \qquad  \Prob(z_i=g| x_i)= \eta_g (x_i).
\label{eqn:latentmix}
\end{eqnarray}
  This suite of models ranges from a standard mixture of experts regression model (in which all model parameters are functions of covariates) to the special cases where some of the model parameters do not depend on covariates. The four models in the ME framework have the following interpretations, see also
  Figure~\ref{fig:MEgraphical}:

\begin{enumerate}[(a)]
\item Mixture models, where the outcome variable distribution depends on the latent cluster membership, denoted $z$. The model is independent of the covariates $x$;
    i.e. $p(y_i,z_i|x_i)= f_{z_i}(y_i | \theta_{z_i}) \eta_{z_i}$.
\item Mixtures of regression models,\index{mixtures!of regression models} where the outcome variable distribution depends on both the covariates $x$ and the latent cluster membership variable $z$; the distribution of the latent variable is independent of the covariates;
    i.e. $p(y_i,z_i|x_i)= f_{z_i}(y_i | \theta_{z_i} (x_i)) \eta_{z_i}$.
\item Simple mixtures of experts models, where the outcome variable distribution depends on the latent cluster membership variable $z$ and the distribution of the latent variable $z$  depends on the covariates $x$;
      i.e. $p(y_i,z_i|x_i)= f_{z_i}(y_i | \theta_{z_i})\eta_{z_i} (x_i) $.
\item Standard mixtures of experts regression models, where  the outcome variable distribution depends on both the covariates $x$ and on the latent cluster membership variable $z$. Additionally the distribution of the latent variable $z$ depends on the covariates $x$;
    i.e. $p(y_i,z_i|x_i)= f_{z_i}(y_i | \theta_{z_i} (x_i)) \eta_{z_i}  (x_i) $.
\end{enumerate}

\noindent The manner in which the different models within the ME framework depend on the covariates is typically application specific. The component weights are usually modelled using a  MNL model, but this need not be the case; \cite{geweke2007} employ a model similar to an ME model, where the component weights have a  multinomial probit structure.\index{probit dependence!multinomial} The form of the distribution $f_g(y_i | \theta_g(x_i))$  depends on the type of outcome data under study. The applications of the ME framework outlined in Section~\ref{sec:apps} include cases where the outcome data range from a  categorical time series, to rank data, to network data.

\section{Statistical Inference for Mixtures of Experts Models}
\label{sec:inference}

Before illustrating the breadth of the ME framework through illustrative applications in Section~\ref{sec:apps}, the issue of inference for ME models is addressed. For any ME model that is underpinned by a finite mixture model, the approaches to inference outlined in Chapter 2 and Chapter 5 in this volume are applicable.
\cite{jacobs91} and \cite{jordan:jacobs:1994} derive maximum likelihood estimates (MLEs) for ME models via the expectation-maximisation (EM) algorithm;\index{EM algorithm!for ME models} \cite{gormley08} employ the closely related expectation-minorisation-maximisation (EMM) algorithm. Estimation of the ME model within the Bayesian framework is detailed, among others, in \cite{peng1996},  \cite{fru-kau:mod}, \cite{vil-etal:reg}, \cite{gormley10}  and  in \citet{fruhwirth2012} in which Markov chain Monte Carlo methods \citep{tanner96} are used; \cite{bishop2002} use variational methods in the Bayesian paradigm to perform inference for a hierarchical mixture of experts model. \cite{hunter2012} present an algorithm for parameter estimation in a semiparametric mixtures of regressions model setting.

 \label{chap13_3}
In this section, a general overview of approaches to inference in the ME framework is provided. Throughout the section, $\ym=(y_1, \ldots, y_n)$ will denote the collection of outcome variables and $\mathbf{x}=(x_1,\ldots,x_n)$ the associated covariates. The latent cluster membership indicators introduced in (\ref{eqn:latentmix}) are denoted by $\zm=(z_1,\ldots,z_n)$, whereas  $\theta = \{\theta_1, \ldots, \theta_G\}$  refers to the collection of the $G$ component parameters  and    $\gamma = \{\gamma_2, \ldots, \gamma_G\}$ to the  unknown parameters in the $G$ component weights.

The exact manner in which an ME model is estimated again depends on the nature of the ME model and the outcome variable. The simple simulated data example of Section~\ref{sec:illustrativeeg} is used here to delineate approaches to inference; more detailed application specific estimation approaches are outlined in Section~\ref{sec:apps}.

\subsection{Maximum likelihood estimation}
\label{sec:mle}

The EM algorithm \citep{dempster77} provides an efficient approach to deriving MLEs in ME models. The EM algorithm is most commonly known as a technique to produce MLEs in settings
where the data under study are incomplete or when optimisation of the likelihood would be simplified if an additional set of variables were known. The iterative EM algorithm consists of an expectation (E) step followed by a maximisation (M) step. Generally, during the\index{EM algorithm!for ME models}
E step the  conditional expectation of the complete (i.e. observed and unobserved) data log likelihood is computed, given the data and current parameter values. In the M step the expected log likelihood is maximised with respect to the model parameters. The imputation of latent variables often makes maximisation of the expected log likelihood more feasible. The parameter estimates produced in the M step are then used in a new E step and the cycle continues until convergence. The parameter estimates produced on convergence are estimates that achieve a stationary point of the likelihood function of the data, which is at least a local maximum but may be a saddle point.

\begin{algorithm}[t!]  \caption{EM algorithm for a simple Gaussian mixture of experts model} \label{gfs:algoEM}
Let $s = 0$. Choose initial estimates for the component weight parameters $\gamma ^{(0)}=(0,\gamma_2^{(0)}, \ldots, \gamma_G^{(0)})$
and for  the component parameters $\mu_{g}^{(0)}$ and $\Sigma_{g}^{(0)}$ for $g = 1, \ldots, G$. 
\begin{algorithmic} \itemsep 0.1cm
\item[1] \textbf{E step}: for $i = 1, \ldots, n$ and $g = 1, \ldots, G$ compute the estimates:
$$ z_{ig}^{(s+1)} = \eta_{g}^{(s)}(x_i | \gamma ^{(s)} ) \phi (y_i | \mu_{g}^{(s)}, \Sigma_{g}^{(s)}) /
\displaystyle \sum_{g'=1}^{G}\eta_{g'}^{(s)}(x_i | \gamma ^{(s)})  \phi (y_i | \mu_{g'}^{(s)}, \Sigma_{g'}^{(s)}) . $$

\item[2] {\bf M step:} Substituting the $z_{ig}^{(s+1)}$ values obtained in the E step
into the log of the complete data likelihood (\ref{eqn:clike}) forms the so called `Q function'
\begin{eqnarray*}
\label{eqn:Q}
Q & = & \sum_{i=1}^{n} \sum_{g=1}^{G} z_{ig}^{(s+1)}
\left[ \tilde{x}_i \gamma_{g} - \log\left\{\sum_{g'=1}^{G} \exp( \tilde{x}_i \gamma_{g'}) \right\} \right.\\
 & & \left. - d/2 \log (2\pi) - 1/2 \log | \Sigma_g| - 1/2(y_i - \mu_g)^\top \Sigma_{g}^{-1} (y_i - \mu_g)  \vphantom{\sum_{g'=1}^{G}}  \right]
\end{eqnarray*}
with $d=\dim(y_i)$, which is maximised with respect to the model parameters.
\begin{itemize}
  \item[(a)]  The updates of the $g = 1, \ldots, G$ component means and covariances are, respectively:
\begin{eqnarray*}
\mu_{g}^{(s+1)} & = & \sum_{i=1}^{n} z_{ig}^{(s+1)} y_{i}/ \displaystyle \sum_{i=1}^{n} z_{ig}^{(s+1)}\\
\Sigma_{g}^{(s+1)} & = & \sum_{i=1}^{n} z_{ig}^{(s+1)} (y_i - \mu_{g}^{(s+1)}) (y_i - \mu_{g}^{(s+1)})^\top / \displaystyle \sum_{i=1}^{n} z_{ig}^{(s+1)}.
\end{eqnarray*}
  \item[(b)] The update for the component weight parameters is obtained via a numerical optimisation step, such as a Newton-Raphson step, where for $g = 2, \ldots, G$
\begin{eqnarray*}
\gamma_{g}^{(s+1)} & = & \gamma_{g}^{(s)} - (H(\gamma_{g}^{(s)}))^{-1} Q^{'}(\gamma_{g}^{(s)})
\end{eqnarray*}
and  $Q^{'}$ and $H$ denote the first and second derivatives of $Q$ with respect to $\gamma_g$ respectively.   Note that this M step is equivalent to fitting a generalised linear model with weights provided by the E step.
\end{itemize}
\item[3] If converged, stop. Otherwise, increment $s$ and return to Step 1.
\end{algorithmic}
\end{algorithm}

The component weights of the simple mixture of experts model outlined in Section~\ref{sec:illustrativeeg} are given by
\begin{eqnarray} \label{MNLME}
\eta_{g}(x_i | \gamma) = \exp\left(\tilde{x}_i \gamma_{g} \right) / \displaystyle \sum_{g'=1}^{G}\exp\left(\tilde{x}_i \gamma_{g'} \right)
\end{eqnarray}
where $\tilde{x}_i = (1, x_i)$ and $\gamma_{g} = (\gamma_{g0}, \gamma_{g1})^\top$. Note that this is a special case of the multinomial logit model. 
For the Normal  distribution $\theta_g = \{\mu_g, \Sigma_g\}$ and the likelihood function of the simple mixture of experts model is
\begin{eqnarray*}
 \lik (\gamma , \theta;G)  
& = & p(\ym | \mathbf{x}, \gamma , \theta)  =  \prod_{i=1}^{n} \displaystyle \sum_{g=1}^{G} \eta_{g}(x_i | \gamma ) \phi( y_i| \mu_{g}, \Sigma_g)  , 
\end{eqnarray*}
 where  $\phi( y_i| \mu_{g}, \Sigma_g)$ is the pdf of the $d$-variate Normal  distribution and  $d=\dim(y_i)$.
It is difficult to directly obtain MLEs from this likelihood. To alleviate this, the data are augmented by imputing for each observation $y_i, i= 1, \ldots, n$,  the latent group membership indicator  $z_i$.  For the EM algorithm, this  latent  variable is represented through  $G$ binary variables
$(z_{i1}, \ldots, z_{iG})$ where $z_{ig}=\indic{z_i=g}$ takes the value 1 if observation $y_i$ is a  member of component $g$ 
and the value 0 otherwise. This provides the complete data likelihood
\begin{eqnarray}
\likc (\gamma , \theta , \zm;G)  
 \!\!\!\! & = & \!\!  p( \ym, \zm | \mathbf{x}, \gamma , \theta) = \prod_{i=1}^{n} \prod_{g=1}^{G} \left\{\eta_{g}(x_i | \gamma ) \phi ( y_i | \mu_{g}, \Sigma_g) \right\}^{z_{ig}}, 
\label{eqn:clike}
\end{eqnarray}
the expectation of (the log of) which is obtained in the E step of the EM algorithm. As the complete data log likelihood is linear in the latent variable, the E step simply consists of replacing for each $i=1,\ldots,n$  the missing data $ z_i$ 
with their expected values  $  \hat{z}_i.$  
In the M step the complete data log likelihood, computed with the estimates  $\hat{\mathbf{z}}=( \hat{z}_1, \ldots,  \hat{z}_n)$, is maximised to provide estimates of the component weight parameters $\hat{\gamma }$ and the component parameters $\hat{\theta}$.

The EM algorithm for fitting ME models is straightforward in principle, but the M step is often difficult in practice. This is usually due to a complex component density and/or component weights model, or a large parameter set. A modified version of the EM algorithm, the Expectation and\index{EM algorithm!for ME models}\index{ECM algorithm}
Conditional Maximisation (ECM) algorithm \citep{meng:rubin:1993} is therefore often employed. In the ECM algorithm, the M step consists of a series of conditional maximisation steps. In the context of the simple mixture of experts example considered here, these maximisations are not straightforward with regard to the $\gamma$ parameters; as in any  MNL  model, no closed form expression for the parameter MLEs is available. Thus, while the conditional M steps for $\mu_g$ and $\Sigma_g$ $\forall g = 1, \ldots, G$ are available in closed form, the conditional M step for $\gamma$ requires the use of a numerical optimisation technique, or as in \cite{gormley08b} the MM algorithm \citep{hunter04a}  in which a minorising function is iteratively maximised and updated.
In summary, to fit the simple mixture of experts example outlined in Section~\ref{sec:illustrativeeg} the EM algorithm proceeds as described in
Algorithm~\ref{gfs:algoEM}.  In the simulated data example, $d=2$.

\cite{mclachlan:peel:2000} outline a number of approaches to assessing convergence in Step~3; typically it is assessed by tracking the change in the log likelihood as the algorithm proceeds.
Standard errors of the resulting parameter estimates are not automatically produced by the EM algorithm, but they can be approximately computed after convergence, for example, by computing and inverting the observed information matrix \citep{mclachlan:peel:2000}.
For a detailed discussion of EM algorithms  in a mixture context  see Chapter 2 and 3 of this volume.

\subsection{Bayesian estimation}
\label{sec:bayesian}

Estimation of ME models can be achieved within the Bayesian paradigm, either using a Markov chain Monte Carlo (MCMC) algorithm\index{MCMC} or via a variational approach.\index{Bayesian estimation!variational} The reader is directed to \cite{bishop2002} for details on the variational approach; this section focuses on inference using MCMC methods.  Both the Gibbs sampler \citep{geman:1984} and the Metropolis-Hastings algorithm \citep{chib95, metropolis:rosenbluth:teller:1953} are typically required. Again, the specific MCMC algorithm, and the form of the prior distributions, depend on the nature of the ME model under study and on the type of the response data.  As is standard in Bayesian estimation of mixture models \citep{dieb:robe:1994, hurn:justel:robert:2003} fitting ME models is greatly simplified by augmenting the observed data with the latent group indicator variable $z_i$ for each observation $y_i$.

\begin{algorithm}[t!]  \caption{MH-within-Gibbs MCMC inference for a simple Gaussian mixture of experts model} \label{gfs:algo}
Iterate the following steps for $m=1,\ldots,M$:
\begin{algorithmic} \itemsep 0.1cm
\item[1] For $g = 1, \ldots, G$, draw $\mu_g$ from the $d$-variate normal posterior  $\Normal (\mu_{ng}, \Lambda_{ng})$ where
\\ $\Lambda_{ng}  =  (\Lambda_{0}^{-1} + n_{g} \Sigma_{g}^{-1})^{-1}$ and
$\mu_{ng}  =  \Lambda_{ng}(\Lambda_{0}^{-1} \mu_{0} +  \Sigma_{g}^{-1}n_g \bar{y}_{g})$
and $n_{g} = \sum_{i=1}^n  \mathbb{I}(z_i = g)$ and $n_{g} \bar{y}_g =  \sum_{i=1}^{n} y_{i}  \mathbb{I}(z_i = g)$.

\item[2] For $g = 1, \ldots, G$, draw $\Sigma_g$ from $\Wishartinv(\nu_{ng}, S_{ng})$ where  $\nu_{ng}  =  \nu_{0} + n_{g}$ and $S_{ng}  =  S_0 + \sum_{i=1}^{n}  \mathbb{I} (z_i = g )(y_i - \mu_{g})(y_i - \mu_{g})^\top.$

\item[3] For $i = 1, \ldots, n$ draw $z_i$ from a multinomial distribution $\Mulnom (1,p_{i1}, \ldots, p_{iG})$   with success probabilities  $(p_{i1}, \ldots, p_{iG})$ where
$$p_{ig} = \eta_{g}(x_i | \gamma ) \phi (y_i| \mu_g, \Sigma_g) / \sum_{g'=1}^{G} \eta_{g'}(x_i|\gamma ) \phi (y_i| \mu_{g'}, \Sigma_{g'}). $$

\item[4] For $g = 2, \ldots, G$, the component weight parameters $\gamma_g$ are updated via a Metropolis-Hastings step,  while holding the
remaining component weight parameters $\gamma _{-g}$ fixed.  Typically,
a multivariate normal proposal distribution $q(\gamma_{g}^{*}| \gamma_g, \gamma _{-g})$ is employed:
\begin{enumerate}
\item[(a)] Propose $\gamma_{g}^{*}  \sim \Normal (\tilde{\mu}_{\gamma}, \tilde{\Lambda}_{\gamma})$ from a  $(\Lold+1)$-variate Normal  distribution
where $\tilde{\mu}_{\gamma}$ and $\tilde{\Lambda}_{\gamma}$ are user specified and might depend on the current value of  $\gamma $.
\item[(b)] If $U \sim U[0,1]$ is such that
$$U \le \min \left\{\frac{
p (\zm  | \gamma_{g}^{*}, \gamma _{-g}, \mathbf{x} ) 
p(\gamma_{g}^{*}) q(\gamma_{g} \: | \: \gamma_g^{*}, \gamma _{-g})}
{p(\zm  | \gamma_{g}, \gamma _{-g}, \mathbf{x} ) 
p(\gamma_{g}) q(\gamma_{g}^{*} \: | \: \gamma_g, \gamma _{-g})} , \: 1 \right\},$$
then set $\gamma_{g} = \gamma _{g}^{*}$; otherwise leave  $\gamma_{g}$ unchanged.
\end{enumerate}
\end{algorithmic}
\end{algorithm}

Performing inference on the illustrative simple mixture of experts model of Section~\ref{sec:illustrativeeg} is again straightforward in principle, but can be difficult in practice. To begin, priors for the model parameters $\mu_g$, $\Sigma_g$ for $g = 1, \ldots, G$ and $\gamma_g$ ($g = 2, \ldots, G$) require specification.  Positing a conditional $d$-variate normal prior  $\Normal (\mu_{0}, \Lambda_{0})$ on the group means $\mu_g$, and an inverse Wishart prior $\Wishartinv(\nu_0, S_0)$ on the group covariances $\Sigma_g$  provides conjugacy for these parameters \citep{hoff:2009}. The full conditional distributions for these parameters are therefore available in closed form, and thus Gibbs sampling can be used to draw samples.

A $({\Lold+1})$-variate normal  $ \Normal (\mu_{\gamma}, \Lambda_{\gamma})$  is an intuitive prior for the component weight parameters $\gamma_g$, but it is non-conjugate. Hence the full conditional distribution is not available in closed form and a Metropolis-Hastings (MH) step can be applied to sample the component weight parameters.
One sweep of such a  Metropolis-within-Gibbs sampler required to fit the simple mixture of experts model of Section~\ref{sec:illustrativeeg} in a Bayesian framework is outlined below. Note that the full conditional distribution of the latent indicator variable $z_i$ for $i = 1, \ldots, n$ is also available in closed form and thus a Gibbs step is available,  see Algorithm~\ref{gfs:algo}.

Sampling the component weight parameters in Step~4 through a MH-algorithm brings  issues such as choosing suitable proposal distributions
$q(\gamma_{g}^{*} \: | \: \gamma_g, \gamma _{-g})$ and  tuning parameters, which may make fitting ME models troublesome. \cite{gormley10b} detail an approach to deriving proposal distributions with attractive properties, within the context of an ME model for network data.

Alternatively, \citet{fruhwirth2012} exploit data augmentation\index{data augmentation} of  the MNL model (\ref{MNLME})  based on the differenced random utility model representation in the context of ME models to implement Step~4.
As shown by \cite{fru-fru:dat},   for each $g=1,\ldots,G$  the MNL model has the following representation as a
binary  logit model conditional on knowing  $\lambda_{hi}=\exp{(\tilde{x}_i \gamma_{h})}$ for all $h \neq g$:
\begin{eqnarray} \label{eq:DAdRUM}
&& u_{gi} = \tilde{x}_i \gamma_{g} - \log(\sum_{h\neq g} \lambda_{hi}) +
\varepsilon_{gi},\\
&& D_i ^g =    \mathbb{I} (u_{gi} \geq 0 )      \nonumber
\end{eqnarray}
where  $u_{gi}$ is a latent variable, $\varepsilon_{gi}$ are   i.i.d.~errors following a logistic distribution,  and $D_i ^g = \mathbb{I} (z_i=g) $ is a binary
outcome variable indicating whether the group indicator $z_i$ is equal to $g$.
Note that $\gamma_{1}=0$ for the baseline, hence $\lambda_{1i}=1$.
In a  data augmented implementation of Step~4, the latent variables   $(u_{2i}, \ldots, u_{Gi})$  are introduced for each $i=1,\ldots,n$ as unknowns.
  Given  $\lambda_{2i}, \ldots, \lambda_{Gi}$ and $z_i$,  $(u_{2i}, \ldots, u_{Gi})$  can be sampled in closed form from exponentially distributed random variables.
Following \citet{sco:dat},  natural proposal distributions  are available to implement an MH-step  to sample $\gamma_{g} |  \gamma _{-g},\zm,\um_g$ for all $g=2,\ldots,G$  conditional on   $\um_g=\{u_{g1}, \ldots, u_{gn}\}$  
from the linear, non-Gaussian regression model  (\ref{eq:DAdRUM}).

To avoid any MH-step,  \citet{fruhwirth2012}  apply auxiliary mixture sampling as introduced by \citet{fru-fru:dat}
to  (\ref{eq:DAdRUM})  and approximate  for  each  $\varepsilon_{gi}$
the logistic distribution by a 10-component scale mixture of Normal  distributions with zero means and parameters $(s_r^2, w_r)$, $r=1, \ldots,10$. In a second step of data augmentation, the component indicator  $r_{gi}$ is introduced as yet another latent variable.
Conditional on the latent variables $\um_g$ and the indicators $\rmm_g =  \{ r_{g1},\ldots,r_{gn}\} $ the binary logit model
(\ref{eq:DAdRUM})  reduces to a linear Gaussian regression model. Hence,   the posterior
$\gamma_{g} |  \gamma _{-g}, \zm,\um_g,\rmm_g$  is Gaussian and a Gibbs step is available to sample $\gamma_{g}$ for all $g=2,\ldots,G$
conditional on $\um_g$ and $\rmm_g$.   Finally, each component indicator $r_{gi}$ is sampled from a discrete distribution conditional on $u_{gi}$  and
$\gamma $.

 Chapter 13  in this volume details Bayesian estimation of informative regime switching models which can be regarded as an  extension of ME models to hidden Markov models in  time series analysis.

As in any mixture model setting, the so called label switching\index{label switching}\index{algorithm!Metropolis-within-Gibbs} problem \citep{stephens:2000, fruhwirth2011} must be considered when employing  such Gibbs based algorithms, see Chapter 5. This identifiability issue, along with others, is discussed in  Section~\ref{sec:identifiability}.

\subsection{Model selection}
\label{sec:modselection}

Within the suite of ME models outlined in Section~\ref{sec:allMEmodels} the question of which, how and where covariates are used naturally arises. This is a challenging problem as the space of ME models is potentially very large, once variable selection for the covariates entering the component weights and the mixture components is considered. Thus in practice only models where covariates enter all mixture components  and/or all component weights as main effects are typically considered in order to restrict the size of the model search space. In fact, even for this reduced model space, there are a maximum of  $G\times 2^\Lold\times2^\Lold$ possible models to consider. In ME models involving generalised linear models of covariates, standard variable selection approaches can be used to find the optimal model. Practical approaches to this issue are detailed in the illustrative applications of Section~\ref{sec:apps}. Note that the manner in which covariates enter the ME model may also be guided by the question of interest in the application
under study.

If the number of components $G$ is unknown, the model search space increases again. Approaches such as marginal likelihood evaluation, or information criteria, are useful for choosing the optimal $G$ in ME models; the reader is referred to  Chapter 7 in this volume which addresses model selection and selecting the number of components in a mixture model in great detail.

\paragraph*{Marginal likelihood computation for mixtures of experts models}
$\:$\\
$\:$\\
As discussed in
Chapter 7, Section 7.2.3.2, highly accurate sampling-based  approximations  to the marginal likelihood  are available, if $G$ is not too large.  For instance,  \citet{fru-kau:mod} apply bridge sampling \citep{fruhwirth:2004} to compute marginal likelihoods  for a  mixture of experts model with a single covariate (that is $q=1$) with up to  four components.
  \citet{fru:pan} combines auxiliary mixture sampling \citep{fru-wag:mar}   with  importance sampling   to   compute marginal likelihoods  for  mixture of experts models. A detailed
  summary of this approach is provided below.

Permutation sampling is applied to ensure that all equivalent modes of the posterior distribution are visited.
 Consider a  permutation $\perm  \in \mathfrak{S}(G)$, where  $\mathfrak{S}(G)$ denotes  the set of the $G!$ permutations of $\{1,\ldots,G\}$.
  To relabel all parameters in a mixture of experts model according to the permutation $\perm$,   define
$\idestar{\theta}_g= \theta_{\perm(g)}$ and  $\idestar{\eta}_g (\Xbetatilde_i) =\eta_{\perm(g)} ( \Xbetatilde_i ) $  for $g=1,\ldots, G$.
Special attention has to be given to the correct relabelling of the coefficients $\gamma_g$ in the  MNL model when applying the permutation $\perm$.
The  coefficients  $(\gamma_1,  \ldots,  \gamma_G)$   and   $(\idestar{\gamma}_1,  \ldots, \idestar{\gamma}_G)$  defining, respectively,  the MNL models
 $ \eta_g ( \Xbetatilde_i)$  and $ \idestar{\eta}_g ( \Xbetatilde_i)$  are related through:
\begin{eqnarray*} 
\Xbetatilde_i  \idestar{\gamma}_{g} &=&    \log \left[ \frac{\idestar{\eta}_g ( \Xbetatilde_i)}{\idestar{\eta}_{g_0} ( \Xbetatilde_i)} \right]  =
 \log \left[ \frac{ {\eta}_{\perm(g)}( \Xbetatilde_i)}{{\eta}_{\perm (g_0)}( \Xbetatilde_i)} \right] =
 \log \left[ \frac{ {\eta}_{\perm (g)} ( \Xbetatilde_i)}{{\eta}_{g_0} ( \Xbetatilde_i)}\right]  -  \log \left[\frac{{\eta}_{\perm (g_0)}( \Xbetatilde_i)}{ {\eta}_{g_0} ( \Xbetatilde_i)} \right] \\
 &=& \Xbetatilde_i  (\gamma_{\perm (g)} - \gamma_{\perm (g_0)}). 
\end{eqnarray*}
To ensure that the baseline  $g_0$ (assumed to be equal to  $g_0=1$   throughout this chapter) remains  the same, despite relabeling,
 the coefficients are permuted in the following  way:
\begin{eqnarray*}  \label{MNLcoo}
\idestar{\gamma}_{g} =    \gamma_{\perm (g)} - \gamma_{\perm (g_0)},  \quad g=1,\ldots, G,
\end{eqnarray*}
which indeed implies that  $\idestar{\gamma}_{g_0}=0$. For $G=2$, the sign of  all coefficients of $\gamma_{2}$ is simply flipped,
if  $\perm=(2,1)$ and remains unchanged, otherwise.

\citet{fru-wag:mar} discuss various importance sampling estimators
of the marginal likelihood   for non-{G}aussian models such as logistic models. Using auxiliary mixture sampling,
one of their approaches constructs the  importance density from the Gaussian full conditional densities
appearing in the augmented Gibbs sampler.   This approach is easily extended to mixture of experts models.
   As discussed in Section~\ref{sec:bayesian},
 auxiliary  mixture sampling yields
 Gaussian  posteriors   $p(\gamma_{g} |  \gamma _{-g}, \zm,\um_g,\rmm_g ) $ for the MNL coefficients  $\gamma_{g}$ in a mixture of experts models,
 conditional on
the latent utilities  $\um_g$ and the latent indicators $\rmm_g$.
 This  allows construction of  an  importance density $q_G (\thmod)$   as in Chapter 7, Section 7.2.3.2,
 however  it is essential  that  $q_G(\thmod)$ covers all symmetric modes of the mixture posterior.
 A successful strategy is to apply random permutation sampling, where each  sampling step is concluded by relabelling as described above,
using a randomly selected permutation  $\perm  \in \mathfrak{S}(G)$. The corresponding importance density reads:
\begin{eqnarray}
\displaystyle  q_G (\thmod)= \frac{1}{S}  \sum_{s=1}^S  \prod_{g=2}^G  p (\gamma_g| \gamma _{-g}  \im{s},  \um_g  \im{s},\rmm_g  \im{s}, \zm  \im{s})
\prod_{g=1}^G   p(\theta_{g}|\zm  \im{s},\ym) ,  \label{QMIXME}
\end{eqnarray}
where  $\{\gamma  \im{s},  \um_2 \im{s},\ldots, \um_G  \im{s},  \rmm_2  \im{s}, \ldots, \rmm_G  \im{s}, \zm  \im{s}\}$, $s=1,\ldots, S$ is a subsequence of posterior draws.
Only if $S$  is large compared to  $G!$, then all symmetric modes are covered by random permutation sampling,
with the number of visits per mode  being on average $S/G!$.
The construction of this importance density is fully automatic and  it is sufficient to store the moments of the various conditional densities (rather than  the
allocations $\zm$ and the latent utilities $\um_g$ and indicators  $\rmm_g$ themselves)
during MCMC sampling  for later evaluation.  This  importance density is used  to compute importance sampling estimators of the marginal likelihood, see the illustrative application in Section~\ref{sec:regressioneg}.



\section{Illustrative Applications}
\label{sec:apps}

The utility of ME models is illustrated in this section through the use of several applications. ME Markov chain models for categorical  time series, ME models for ranked preference data, and ME models for network data, all of which are members of the ME model framework, are applied.

\subsection{Analysing marijuana use  through ME  Markov chain models}
\label{sec:regressioneg}

\cite{lan-etal:ass} studied data on the marijuana use\index{marijuana use} of  237  teenagers taken from five annual waves (1976-80) of the National Youth Survey. The  respondents  were 13 years old in 1976  and reported for five consecutive years their marijuana use in the past year as a categorical variable with the three categories \lq\lq never\rq\rq , \lq\lq not more than once a month\rq\rq\ and  \lq\lq more than once a month\rq\rq .  Hence,  for $i = 1, \ldots,  237$,  the outcome variable is a categorical  time series $y_i=(y_{i0},  y_{i1}, \ldots, y_{i4} )$   with   three states, labeled  1 for  never-user,    2 for light and  3 for
heavy users.

 To identify groups of  teenagers with similar marijuana use behaviour,  \citet{fru:pan}  applied a ME approach based on Markov chain models\index{Markov chain models!ME} \citep{fruhwirth2012} and considered
each time series $y_i$  as a single entity  belonging to one of  $G$ underlying classes.
  Various types  of ME  Markov chain models   were applied  to capture dependence in marijuana use over time and
  to investigate if   the  gender  of the teenagers  can be associated with a certain type of marijuana use.

Given the times series nature of the categorical outcome variable $y_i$,   the component density $f_g(\cdot)$ in the mixture of experts model (\ref{eqn:genericME}) must have an appropriate form and various models are considered.
Model ${\cal M}_1$ is a  standard finite mixture of  time-homogeneous Markov chain models of order one \citep{pam-fru:mod}
 where each  component-specific density $f_g(\cdot )$ in (\ref{eqn:genericME}) is characterized by
 a   transition matrix  $\xiv_g$ with $J=3$  rows  and the weight distribution $\eta_1,\ldots, \eta_G$ is independent of any covariates.  Each  row $\xiv_{g,j\cdot} = (\xi_{g,j1}, \ldots, \xi_{g,j3})$, $j=1,\ldots, J$, of the  matrix  $\xiv_g$ represents a probability distribution over the three categories of marijuana use with
\begin{eqnarray*}
 \xi_{g,jk} = \Prob(y_{it} = k | y_{i,t-1}=j,z_i=g), \qquad k=1,\ldots,3. 
\end{eqnarray*}
This model is extended in various ways to include covariate information into the transition behaviour.
First, an  inhomogeneous model (labelled model ${\cal M}_2$) is considered, where the transition matrix in each group depends on the gender $x_i$ of  the teenager. If  all $J=6$ possible combinations $\hist_{it}=(y_{i,t-1},x_{i})$  of  the immediate  past $y_{i,t-1}$ at time $t$ and the gender $x_i$ are indexed by $j=1,\ldots,J$, then the component-specific density  $f_g(y_i|\xiv_g)$ in (\ref{eqn:genericME})   can be described by  a generalized transition matrix $\xiv_g$ with six rows, with the $j$th row  $ \xiv_{g,j \cdot}=(\xi_{g,j1}, \ldots, \xi_{g,j3})$ describing again the conditional distribution of  $y_{i t}$, given that the state of  the history $\hist_{it}$ equals $j$:
 \begin{eqnarray*} \label{mnltrreg}
 \xi_{g,jk}=\Prob(y_{i t}= k|\hist_{it}=j , z_i=g ),  \quad  k=1,\ldots,3.
\end{eqnarray*}
Evidently,  the component specific distribution  reads:
\begin{eqnarray}
f_g(y_i|\xiv_g)= \prod_{j=1}^J \prod_{k=1}^3
\xi_{g,jk}^{n_{i,jk}} \label{modculgtr}
\end{eqnarray}
where,  for each  time series $i$,
$ n_{i,jk}= \sum_{t=1}^4  \mathbb{I} (y_{it}= k, \hist_{it}= j )$   
is the number of transitions  into state $k$ given a history of type $j$.
 Note that   (\ref{modculgtr})  is formulated conditional on the first observation $y_{i0}$.

Alternative  component-specific distributions can be constructed,  by  defining the history  $\hist_{it}  $ through  different combinations of
past values and covariates. Choosing  $\hist_{it}=(y_{i,t-1},t)$, for instance,
defines a  time-inhomogeneous Markov\index{Markov chain models!inhomogeneous}  chain model, labeled model ${\cal M}_3$,  with $J=12$ different covariate combinations.
This model is able to  capture the effect that
the transition behaviour between the states might change  as the teenagers grow older.

The most complex model, labelled model ${\cal M}_4$,   extends model ${\cal M}_3$ by assuming additional dependence on gender,
  i.e.~$ \hist_{it}=(y_{i,t-1}, t, x_i )$, with $J=24$ different covariate combinations.
Both model ${\cal M}_3$ and ${\cal M}_4$ are  characterised by component-specific
generalized transition matrices $\xiv_g$ with, respectively, 12 and 24 rows.
For each of  the models  ${\cal M}_2, {\cal M}_3, {\cal M}_4$,  it is assumed that the  weight distribution $\eta_1,\ldots, \eta_G$  is independent of any covariate,  leading to various finite mixtures of  inhomogeneous Markov chain models.

\begin{table}[t!]
\begin{center}
\caption{Marijuana data; marginal likelihood $\log p(\ym|{\cal M}_k)$ for various finite mixtures of  homogeneous (${\cal M}_1$) and inhomogeneous (${\cal M}_2,   {\cal M}_3, {\cal M}_4$) Markov chain models  with an increasing number
 $G$ of classes (best values for each model in bold font) \label{tab2}}
 \begin{tabular}{cccccc}
  & &  & \multicolumn{3}{c}{$G$}\\ 
  Model & Covariates & $J$ & 1 & 2& 3\\
   \hline
   ${\cal M}_1$ &    - &  3& -605.5 &      {\bf -600.0}     &   -600.3   \\
${\cal M}_2$ & $ x_i $  & 6 &  -610.0     &     {\bf -601.3}     & -603.6   \\
 ${\cal M}_3$ & $t$   & 12 &  -613.7    &  {\bf  -596.5}     &  -599.4    \\
 ${\cal M}_4$ & $ t, x_i $  &  24 & -619.8    &  -602.7   &{\bf -601.1} \\
\end{tabular}
\end{center}
\end{table}

  Bayesian inference is carried out  for all models ${\cal M}_1, \ldots, {\cal M}_4$   for  an increasing  number  $G=1,2, 3$ of classes.
 MCMC estimation  as described in Section~\ref{sec:bayesian} is easily applied, as  the $J$ rows $\xiv_{g, j \cdot}$ of $ \xiv_{g}$    are conditionally independent  under  the conditionally conjugate Dirichlet prior $ \xiv_{g, j \cdot} \sim  \Dir (d_{0,j1},\ldots,d_{0,j3} )$. Given $\zm$ and $\ym $, 
 the generalized transition matrix $ \xiv_{g}$ is sampled row-by-row from a total of $JG$ Dirichlet distributions:
  \begin{eqnarray}  \label{postxi}
  \xiv_{g, j \cdot}|\zm,\ym \sim \Dir (d_{0,j1} +  n^g_{j1},\ldots,d_{0,j3} +  n^g_{j3}),  \quad j=1,\ldots,J,\ g=1,\ldots,G,
\end{eqnarray}
where $n^g_{jk} =  \sum_{i:z_i=g} n_{i,jk}$ is the total number of
transitions into state $k$ observed in class  $g$ given a history of type $j$.

 For model comparison, the marginal likelihood\index{marginal likelihood} is computed explicitly for $G=1$,
while importance sampling as described in Section~\ref{sec:modselection} is applied for $G=2, 3$,
 using   the importance density:
\begin{eqnarray*}
\displaystyle  q_G (\thmod)
= \frac{1}{S}  \sum_{s=1}^S  p (\eta|   \zm  \im{s})  \prod_{g=1}^G  \prod_{ j=1} ^J  p(\xiv_{g, j \cdot}|\zm  \im{s},\ym)  ,
\end{eqnarray*}
where   $p(\xiv_{g, j \cdot}|\zm ,\ym)$ is equal to the full conditional Dirichlet posterior of  $\xiv_{g, j \cdot}$ given in (\ref{postxi}).
Random permutation sampling is applied to ensure that all $G!$ symmetric modes are visited and $S=10,000$.
 The marginal likelihoods  reported in Table~\ref{tab2} select $G=2$ for all   models  except
   for    ${\cal M}_4$,  where  $G=3$ is selected.
   Among all models,  the marginal likelihood is the highest for  model ${\cal M}_3$  with $G=2$ classes.

   Hence, a  time-inhomogeneous Markov chain model which does not depend on gender  best describes the transition behaviour in each class.
   Table~\ref{tabest1} reports the corresponding posterior means $\Ew(\xiv_{g,\cdot}|\ym)$ and $\Ew(\eta_{g}|\ym)$ for each of the two groups.
  Label switching was resolved by applying $k$-means clustering to  a vector constructed from  all persistence probabilities at all time points.
Both groups are roughly of equal size, with the first group being slightly larger. A characteristic difference is evident for the two groups of teenagers.
In group 1, never-users have a high probability $\xi_{t,11}$ to remain  never-users throughout the whole observation period, whereas this probability is much smaller for the second group right from the beginning and  drops to only 45\% in the last year.

\begin{table}[t!]
\begin{center}
\caption{Marijuana data; finite mixture of time-inhomogeneous Markov chain  models (model ${\cal M}_3$)  with  $G=2$  classes;
the estimated posterior mean  $\Ew(\xiv_{g,\cdot}|\ym)$  is arranged   for each $t=1,\ldots,4$ as a $3\times3$ matrix; the estimated class sizes
 $\hat{\eta}_g$ are equal to the posterior mean $\Ew(\eta_{g}|\ym)$ \label{tabest1}}
 \begin{tabular}{lcccccccccccccc}
 & \multicolumn{3}{c}{$t=1$ } & \multicolumn{3}{c}{$t=2$ } &  \multicolumn{3}{c}{$t=3$ } & \multicolumn{3}{c}{$t=4$ } \\
  \hline
Group 1&  0.93 &  0.04 &  0.03 & 0.89  &  0.09 &  0.02 & 0.90 &  0.04 &   0.06 & 0.93 &  0.04 &  0.03\\
 $(\hat{\eta}_1=0.56)$ & 0.50 &   0.17 &   0.34 &  0.10 &   0.33 &   0.57 & 0.17 &   0.65 &   0.18 & 0.20 &   0.64 &   0.16   \\
& 0.22 &   0.18 &   0.60 & 0.10 &    0.17 &   0.73 & 0.04 &   0.27 &   0.69 & 0.15 &   0.12 &   0.74   \\
  \hline
 Group 2  &  0.76 &  0.21 &  0.03   &   0.70 &  0.24 &  0.06 &   0.75 &  0.18 &   0.07  &   0.45 &  0.43 &   0.12  \\
$(\hat{\eta}_2=  0.44)$  &   0.34 &  0.15 &   0.51 &   0.23 &  0.39 &   0.38 &   0.31 &   0.41 &   0.28  &   0.46 &  0.43 &   0.11\\
 &   0.18 &  0.23 &   0.59&     0.10 &  0.22 &   0.68  &   0.13 &  0.15 &   0.72 & 0.05 &  0.10 &   0.85 \\
\end{tabular}
\end{center}
\end{table}

\begin{table}[t!]
\begin{center}
\caption{Marijuana data; ME model  with  $\tilde{x}_{i}=(1,x_i,D_{i0})$ (model ${\cal M}_5$), extending
model ${\cal M}_3$  with  $G=2$  classes.
 Posterior expectation and  95\% HPD region of  the component weight parameters   $\gamma_{2j}$
 in the ME model (\ref{MNLME}) \label{tabest2}}
\begin{tabular}{lcc}
Covariate $\tilde{x}_{ij}$ & $\Ew(\gamma_{2j}|\ym)$ & \multicolumn{1}{c}{95\% HPD region of  $\gamma_{2j}$}  \\
\hline
constant&    -0.69  &   (-1.75,0.35) \\
male (baseline: female) &  0.28 &  (-0.71,1.22)   \\
marijuana use  in 1976 (baseline: no)& -0.07  &    (-1.70,1.43) \\
 \hline
   $\log p(\ym|{\cal M}_5)$ &  -598.5   &     \\ 
\end{tabular}
\end{center}
\end{table}

To investigate if gender is associated with group membership, model ${\cal M}_3$ with $G=2$ classes is combined
with the  ME model (\ref{MNLME}),  by including  gender  as subject-specific covariate  $x_i$ as in the example in Section~\ref{sec:illustrativeeg}.
This model is labelled model ${\cal M}_5$. Additionally,   a  dummy variable  $ D_{i0}$ is included, indicating if
 the teenager used  marijuana, light or heavy, in the first year.
   As  $G=2$, the ME model (\ref{MNLME}) reduces to a binary logit model  with  regression coefficients $ \gamma_2=(\gamma_{20},\gamma_{21}, \gamma_{22})$, each assumed to follow  a standard normal  prior distribution.

From   posterior inference  in  Table~\ref{tabest2},    we find
that male teenagers have a slightly  higher probability to
belong to the second group, because  $\Ew(\gamma_{21}|\ym)>0$,
however, the coefficient $\gamma_{21}$ is not significantly different from 0. Similarly, the initial
state from which a teenager started in  1976
 does not have a significant influence on the probability to
belong to the second group. 
 This suggests that the ME time-inhomogeneous Markov chain model  actually reduces to a standard mixture of time-inhomogeneous Markov chain models   which
  is confirmed by comparing the log marginal likelihood of both models,
 being  equal to  -596.5   for a standard mixture model with $G=2$ groups, see Table~\ref{tab2},  and being equal to  -598.5
 for an ME model with $G=2$ groups, see  Table~\ref{tabest2}.

The marginal likelihood estimator for the  ME model is based  on importance sampling using
  the importance density (\ref{QMIXME})  derived from auxiliary  mixture sampling:
\begin{eqnarray*}
\displaystyle  q_G (\thmod)
= \frac{1}{S}  \sum_{s=1}^S   p (\gamma_2|  \um_2  \im{s},\rmm_2  \im{s}, \zm  \im{s})
\prod_{g=1}^2  \prod_{ j=1} ^J  p(\xiv_{g, j \cdot}|\zm  \im{s},\ym)  ,
\end{eqnarray*}
where   $p(\gamma_{2} |  \um_2,\rmm_2,\zm) $  is conditionally Gaussian
and $p(\xiv_{g, j \cdot}|\zm ,\ym)$ is equal to the full conditional Dirichlet posterior of  $\xiv_{g, j \cdot}$ given in (\ref{postxi}).
Again, random permutation sampling is applied to ensure that the two equivalent modes are visited.

To sum up,   this  investigation shows that teenagers may, indeed,  be clustered into two groups with different behaviour
 with respect to marijuana use,  one being a never-user group, while the second group  has   a much higher risk to become a user.
Preference for  a  standard mixture of Markov chain models over a mixture of experts Markov chain model  based on  gender
 shows that  the two types of marijuana use   cannot be associated with the gender of the teenager.
Both  male and female teenagers have about the same risk to belong to the second group.
Unobserved factors, not the gender,
 are relevant for membership of  a teenager  to one group or the other.

\subsection{A mixture of experts model for ranked preference data}
\label{sec:rankeg}

Mary McAleese\index{Mary McAleese}\index{ranked preference data} served as the eighth President of Ireland from 1997 to 2011 and was elected under the Single Transferable Vote electoral system.\index{single transferable vote system} Under this system voters rank, in order of their preference, some or all of the electoral candidates. The vote counting system which results in the elimination of candidates and the subsequent election of the President is an intricate process involving the transfer of votes between candidates as specified by the voters' ballots. Details of the electoral system, the counting process and the 1997 Irish presidential election are given in \cite{coakley04}, \cite{sinnott95}, \cite{sinnott99} and \cite{marsh99}.

The 1997 presidential election race involved five candidates: Mary Banotti, Mary McAleese, Derek Nally, Adi Roche and Rosemary Scallon. Derek Nally and Rosemary Scallon were independent candidates while Mary Banotti and Adi Roche were endorsed by the then current opposition parties Fine Gael and Labour respectively. Mary McAleese was endorsed by the Fianna F\'{a}il party who were in power at that time. In terms of candidate type, McAleese and Scallon were deemed to be conservative candidates with the other candidates regarded as liberal. \cite{gormley08, gormley08b, gormley10, gormley10b} provide further details on the 1997 presidential election and on the candidates.

One month prior to election day a survey was conducted by Irish Marketing Surveys on 1083 respondents. Respondents were asked to list some or all of the candidates in order of preference, as if they were voting on the day of the poll.\index{poll} In addition, pollsters gathered data on attributes of the respondents as detailed in Table~\ref{tab:votercovs}.

\begin{table}[t!]
\begin{center}
\caption{Covariates recorded for each respondent in the Irish Marketing Surveys poll. \label{tab:votercovs}}{}
\begin{tabular}{llllll}
\textbf{Age} &  \textbf{Area} & \textbf{Gender} & \textbf{Government} & \textbf{Marital} & \textbf{Social}\\
 & & & \textbf{satisfaction} & \textbf{status} & \textbf{class}\\\hline
-- &  City &Housewife &No opinion &Married &AB\\
 & Rural &Male& Not satisfied &Single &C1\\
 & Town & Non-housewife & Satisfied &Widowed &C2\\
 & & & & &DE\\
 & & & & &F50+\\
 & & & & &F50-\\
\end{tabular}
\end{center}
\end{table}

Interest lies in determining if groups of voters with similar preferences (i.e. voting blocs) exist within the electorate. If such voting blocs do exist, the influence the recorded socio-economic variables may have on the clustering structure and/or on the preferences which characterize a voting bloc is also of interest. Jointly modelling the rank preference votes and the covariates through a mixture of experts model for rank preference data when clustering the electorate provides this insight.

Given the rank nature of the outcome variables or votes $y_i$ ($i = 1, \ldots, n = 1083$) the component density $f_g(\cdot)$ in the mixture of experts model (\ref{eqn:genericME}) must have an appropriate form. The Plackett-Luce model \citep{plackett75, gormley06} (or exploded logit model) for rank data provides a suitable model; Benter's model \citep{benter94} provides another alternative. Let $y_i = [c(i,1), \ldots, c(i, m_i)]$ denote the ranked ballot of voter $i$ where $c(i,j)$ denotes the candidate ranked in $j$th position by voter $i$ and $m_i$ is the number of candidates ranked by voter $i$. Under the Plackett-Luce model, given that voter $i$ is a member of voting bloc $g$ and given the `support parameter' $p_{g} = (p_{g1}, \ldots, p_{gM})$, the probability of voter $i$'s ballot is
\begin{eqnarray*}
 p(y_i | p_g) = \frac{p_{g, c(i,1)}}{\sum_{s=1}^{M} p_{g, c(i,s)}} \cdot \frac{p_{g, c(i,2)}}{\sum_{s=2}^{M} p_{g , c(i,s)}} \cdots \frac{p_{g, c(i,m_{i})}}{\sum_{s=m_{i}}^{M} p_{g, c(i,s)}},
  \end{eqnarray*}
where $M = 5$ denotes the number of candidates in the electoral race. The support parameter $p_{gj}$ (typically restricted such that $\sum_{j=1}^{M} p_{gj} = 1$) can be interpreted as the probability of ranking candidate $j$ first, out of the currently available choice set. Hence, the Plackett-Luce model models the ranking of candidates by a voter as a set of independent choices by the voter, conditional on the cardinality of the choice set being reduced by one after each choice is made.

\begin{table}[t!]
\begin{center}
\caption{The model with smallest BIC within each type of mixture of experts
model for ranked preference data applied to the 1997 Irish presidential election data \label{tab:BICs}}{}
\begin{tabular}{llll}
 & BIC  & $G$ &  Covariates\\\hline
Simple mixture of experts model & 8491 &  4  & $\eta_{g}$: Government satisfaction,  Age. \vspace{0.2cm} \\
Standard mixture of experts  &  8512 & 3 &  $\eta_{g}$: Government satisfaction,  Age.\\
regression model & & & $p_{g}$: Age  \vspace{0.2cm} \\
Mixture model & 8513 & 3& --  \vspace{0.2cm} \\
Mixture of regressions model & 8528 & 1 & $p_{g}$: Government satisfaction\\ 
\end{tabular}
\end{center}
\end{table}

In the standard mixture of experts regression model, the parameters of the component densities are modelled as a function of covariates. Here the support parameters are modelled as a logistic function of the covariates
\begin{eqnarray*}
\log \left[\frac{p_{gj}(x_i)}{p_{g1}(x_i)} \right] = \beta_{gj0} + \beta_{gj1} x_{i1} + \cdots + \beta_{gj\Lold} x_{i\Lold}
  \end{eqnarray*}
where $x_i = (x_{i1}, \ldots, x_{i\Lold})$ is the set of $\Lold$ covariates associated with voter $i$ and $\beta_{gj}=( \beta_{gj0}, \ldots, \beta_{gj\Lold}  )^\top $  are unknown parameters  for $j=2, \ldots, M$. Note that for identifiability reasons candidate 1 is used as the baseline choice and $\beta_{g1} = (0, \ldots, 0)$ for all $g = 1, \ldots, G$.

\begin{figure}[t!]
\begin{center}
\hspace{-2cm} \scalebox{0.7}{\includegraphics[angle=270]{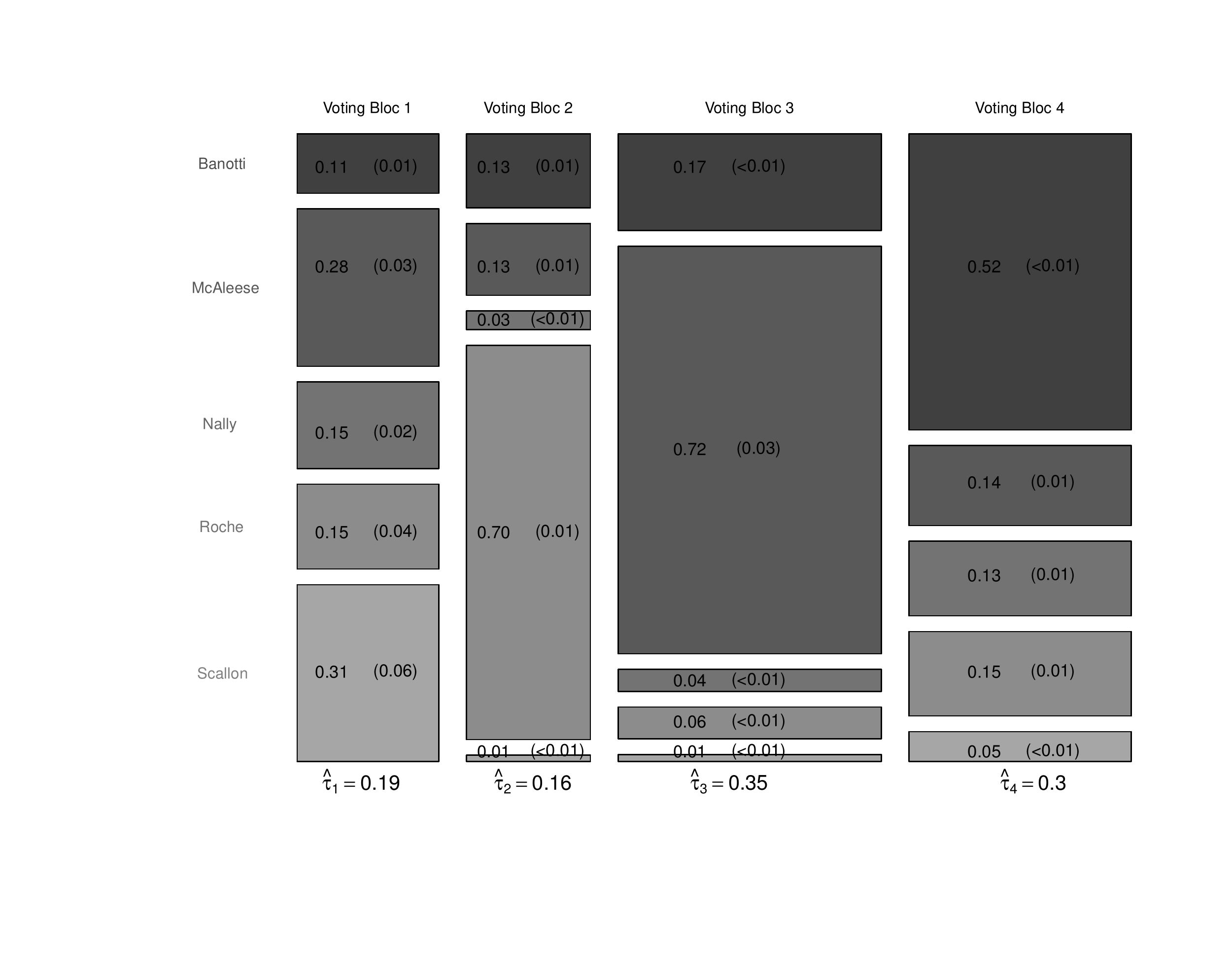}}
\caption{A mosaic plot representation of the parameters of the component densities of the simple mixture of experts model for rank preference data. The width of each block is proportional to the marginal probability of component membership ($\hat{\eta}_g = \sum_{i=1}^{n} \eta_{g}(x_i | \hat{\gamma })/n$). The blocks are divided in proportion to the Plackett-Luce support parameters which are detailed therein. Standard errors are provided in parentheses. \label{fig:mosaic}}{}
\end{center}
\end{figure}

In the standard mixture of experts regression model, the component weights are also modelled as a function of covariates, in a similar vein to the example used in Section~\ref{sec:illustrativeeg}, i.e.
\begin{eqnarray*}
  \log \left[\frac{\eta_{g}(x_i)}{\eta_{1}(x_i)} \right] = \gamma_{g0} + \gamma_{g1}x_{i1}  + \cdots + \gamma_{g\Lold} x_{i\Lold},
\end{eqnarray*}
where voting bloc 1 is used as the baseline voting bloc.

The suite of four ME models in the ME framework (Figure \ref{fig:MEgraphical}) arise from modelling the component parameters and/or the component weights as functions of covariates, or as constant with respect to covariates. In this application, each model is fitted in a maximum likelihood framework using the EM algorithm; approximate standard errors for the model parameters are derived from the empirical
information matrix \citep{mclachlan:peel:2000} after the EM algorithm has converged. Model fitting details for each model are outlined in \cite{gormley08,gormley08b,gormley10, gormley10b}.

Each of the four ME models for rank preference data were fitted to the data from the electorate in the Irish presidential election poll. A range of models with
 $G = 1, \ldots, 5$ was considered and a forward step-wise selection method was employed to choose influential covariates. The Bayesian Information Criterion (BIC) \citep{kass:raftery:1995, schwarz:1978} was used to select the optimal model; this criterion is a penalized likelihood criterion which rewards model fit while penalizing non-parsimonious models, see also Chapter~7, Section~7.2.2 of this volume.  Small BIC values indicate a preferable model. Table \ref{tab:BICs} details the optimal models for each type of ME model fitted.

Based on the BIC values, the optimal model is a simple mixture of experts
model with four groups where  \lq\lq age\lq\lq\ and \lq\lq government satisfaction\rq\rq\  are  important
covariates for determining group or  \lq\lq voting bloc\rq\rq\   membership. Under this simple mixture of experts model, the covariates are not informative within voting blocs, but only in determining voting bloc membership. The maximum likelihood estimates of the model parameters are reported in Figure \ref{fig:mosaic}\index{mosaic plot} and in Table \ref{tab:odds}.

\begin{table}[t!]
\begin{center}
\caption{Odds ratios ($\exp(\gamma_g)/\exp(\gamma_1)$) for the component weight parameters in the  simple
ME model for rank preference data (95\% confidence intervals are given in parentheses). The covariates `age' and `government satisfaction level' were selected as influential \label{tab:odds}}{}
{
\begin{tabular}{cccc}
 & \multicolumn{1}{c}{{\bf Age}} & \multicolumn{1}{c}{{\bf Not satisfied}} & \multicolumn{1}{c}{{\bf Satisfied}}\\\hline
Voting bloc 2& 0.01 (0.00, 0.05) &2.80   (0.77, 10.15) &1.14  (0.42, 3.11) \\
Voting bloc 3 &0.95  (0.32, 2.81) &3.81 (0.90, 16.13) & 3.12  (0.94, 10.31)\\
Voting bloc 4 &1.56  (0.35, 6.91) &3.50  (1.07, 11.43) & 0.35  (0.12, 0.98) \\ 
\end{tabular}}
\end{center}
\end{table}

The support parameter estimates illustrated in Figure \ref{fig:mosaic} have an interpretation in the context of the 1997 Irish presidential election. Voting bloc 1 could be characterized as the  \lq\lq conservative voting bloc\rq\rq\ due   to its large support parameters for McAleese and Scallon. Voting bloc 2 has large support for the liberal candidate Adi Roche. Voting bloc 3 is the largest voting bloc in terms of marginal component weights and intuitively has larger support parameters for the
high profile candidates McAleese and Banotti. These candidates were endorsed by the two largest political parties in the country at that time. Voters belonging to voting bloc 4 favor Banotti and have more uniform levels of support for the other candidates. A detailed discussion of this optimal model is also given in \cite{gormley08b}.

Table \ref{tab:odds} details the odds ratios computed from the component weight parameters $\gamma = \{\gamma_2, \gamma_3, \gamma_{4}\}$. In the model, voting bloc 1 (the conservative voting bloc) is the baseline voting bloc and $\gamma_1 = (0, \ldots, 0)^\top$. Two covariates were selected as influential: age and government satisfaction levels. In the  \lq\lq government satisfaction\rq\rq\ covariate, the baseline was chosen to be  \lq\lq no opinion\lq\lq .

Interpreting the odds ratios provides insight to the type of voter which characterises each voting bloc. For example, older (and generally more conservative) voters are much less likely to belong to the liberal voting bloc 2 than to the conservative voting bloc 1 ($\exp(\gamma_{21}) = 0.01$). Also, voters with some interest in government are more likely to belong to voting bloc 3 ($\exp(\gamma_{32}) = 3.81$ and $\exp(\gamma_{33}) = 3.12$), the bloc favouring candidates backed by large government parties, than to belong to the conservative voting bloc 1. Voting bloc 1 had high levels of support for the independent candidate Scallon. The component weight parameter estimates further indicate that voters dissatisfied with the current government are more likely to belong to voting bloc 4 than to voting bloc 1 ($\exp(\gamma_{42}) = 3.50$). This is again intuitive as voting bloc 4 favours Mary Banotti who was backed by the main government opposition party, while voting bloc 1 favours the government backed Mary McAleese. Further  interpretation of the component weight parameters are given in \cite{gormley08b}.

\subsection{A mixture of experts latent position cluster model}
\label{sec:networkeg}

The latent position cluster model\index{latent position cluster model} \citep{handcock07} develops the idea of the latent social space \citep{hoff02} by extending it to accommodate clusters of actors in the latent space. Under the latent position cluster model, the latent location of each actor is assumed to be drawn from a finite normal mixture model, each component of which represents a cluster of actors. In contrast, the model outlined in \cite{hoff02} assumes that the latent positions were normally distributed. Thus, the latent position cluster model offers a more flexible version of the latent space model for modelling heterogeneous social networks.\index{social networks}

The latent position cluster model provides a framework in which actor covariates may be explicitly included in the model -- the probability of a link between two actors may be modelled as a function of both their separation in the latent space and of their relative covariates. However, the covariates may contribute more to the structure of the network than solely through the link probabilities -- the covariates may influence both the cluster membership of an actor and their link probabilities. A latent position cluster model in which the cluster membership of an actor is modelled as a function of their covariates lies within the mixture of experts framework.

Specifically, social network data take the form of a set of relations $\{y_{i,j}\}$ between a group of $i, j = 1, \ldots, n$ actors, represented by an $n \times n$ sociomatrix $\ym$. Here it is assumed that the relation $y_{i,j}$ between actors $i$ and $j$ is a binary
relation, indicating the presence or absence of a link between the two actors; the mixture of experts latent position cluster model is easily extended to other forms of relation (such as count data). Covariate data $x_i = (x_{i1}, \ldots, x_{i\Lold})$ associated with actor $i$ are assumed to be available, where $\Lold$ denotes the number of observed covariates.

\begin{table}[t!]
\begin{center}
\caption{Covariates associated with the 71 lawyers in the US corporate law firm. The last category in each categorical covariate is treated as the baseline category in all analyses. \label{tab:lawcovs}}
\begin{tabular}{ll} 
\textbf{Covariate} & \textbf{Levels}\\\hline
 Age & -- \\
Gender & 1 = male\\
 & 2 = female\\
Law school &  1 = Harvard or Yale\\
 & 2 = University of Connecticut\\
 & 3 = other\\
Office &  1 = Boston\\
 & 2 = Hartford\\
 & 3 = Providence\\
Practice &  1 = litigation\\
               & 2 = corporate\\
Seniority &  1 = partner\\
 & 2 = associate\\
Years with the firm & --\\ 
\end{tabular}
\end{center}
\end{table}

Each actor $i$ is assumed to have a location $w_i = (w_{i1} , \ldots, w_{i \dold })$ in the $\dold $ dimensional latent social space. The probability of a link between any two actors is assumed to be independent of all other links in the network, given the latent locations of the actors. Let $x_{i,j} = (x_{ij1}, \ldots, x_{ij\Lold})$ denote an $\Lold$ vector of dyadic specific covariates where $x_{ijk} = d(x_{ik}, x_{jk})$ is a measure of the similarity in the value of the $k$th covariate for actors $i$ and $j$. Given the link probabilities parameter vector $\beta$,  the likelihood function is then
\begin{eqnarray*}
p(\ym | \wm, \mathbf{x}, \beta) = \prod_{i=1}^{n} \prod_{j \neq i} p(y_{i,j} | w_i, w_j, x_{i,j}, \beta)
  \end{eqnarray*}
where $\wm$ is the $n \times \dold $ matrix of latent locations and $\mathbf{x}$ is the matrix of dyadic specific covariates. The probability of a link between actors $i$ and $j$ is then modelled using a logistic regression model where both dyadic specific covariates and Euclidean distance in the latent space are covariates:
\begin{eqnarray*}
\log \left[\frac{\Prob(y_{i,j} = 1)}{\Prob (y_{i,j} = 0)} \right] = \beta_0 + \beta_1 x_{ij1} + \cdots + \beta_{\Lold} x_{ij\Lold} - ||w_i - w_j||.
  \end{eqnarray*}
To account for clustering of actor locations in the latent space, it is assumed that the latent locations $w_i$ are drawn from a finite mixture model. Moreover, in the mixture of experts latent position cluster model, the latent locations are assumed drawn from a finite mixture model in which actor covariates may influence the mixing proportions:
\begin{eqnarray*}
w_i \sim \sum_{g=1}^{G} \eta_{g}(x_i | \gamma ) \phi (w_i | \mu_g, \sigma^{2}_{g} I )
  \end{eqnarray*}
where
\begin{eqnarray*}
\eta_{g}(x_i | \gamma ) = \frac{\exp(\gamma_{g0} + \gamma_{g1}x_{i1} + \cdots + \gamma_{g\Lold}x_{i\Lold})}{\sum_{g'=1}^{G} \exp(\gamma_{g'0} + \gamma_{g'1}x_{i1} + \cdots + \gamma_{g'\Lold}x_{i\Lold})}
  \end{eqnarray*}
and $\gamma_{1} = (0, \ldots, 0) ^\top$. This model has an intuitive motivation: the covariates of an actor may influence their cluster membership, their cluster membership influences their latent
location, and in turn their latent location determines their link probabilities.

The mixture of experts latent position cluster model can be fitted within the Bayesian paradigm; as outlined in Section~\ref{sec:bayesian} a Metropolis-within-Gibbs sampler can be employed to draw samples from the posterior distribution of interest. Model issues such as likelihood invariance to distance preserving transformations of the latent space and label switching must be  considered during the model fitting process -- an approach to dealing with such model identifiability\index{identifiability!model} and full model fitting details are available in \cite{gormley10b}.  In this application, model choice concerns not only the number $G$ of clusters, but also the dimension  $\dold $ of the latent space.\index{model choice}

An example of the mixture of experts latent position cluster model methodology is provided here through the analysis of a network data set detailing interactions between a set of 71 lawyers in a corporate law firm in the USA \citep{lazega01}.  The data include measurements of the coworker network, an advice network and a friendship network. Covariates associated with each lawyer in the firm are also included and are detailed in Table \ref{tab:lawcovs}. Interest lies in identifying social processes within the firm such as knowledge sharing and organisational structures, and examining the potential influence of covariates on such processes.

\begin{figure}[t!]
\begin{center}
\begin{tabular}{cc}
\subfigure[Latent position cluster model.]{\label{fig:box1}\includegraphics[width=60mm, height=65mm, angle=270]{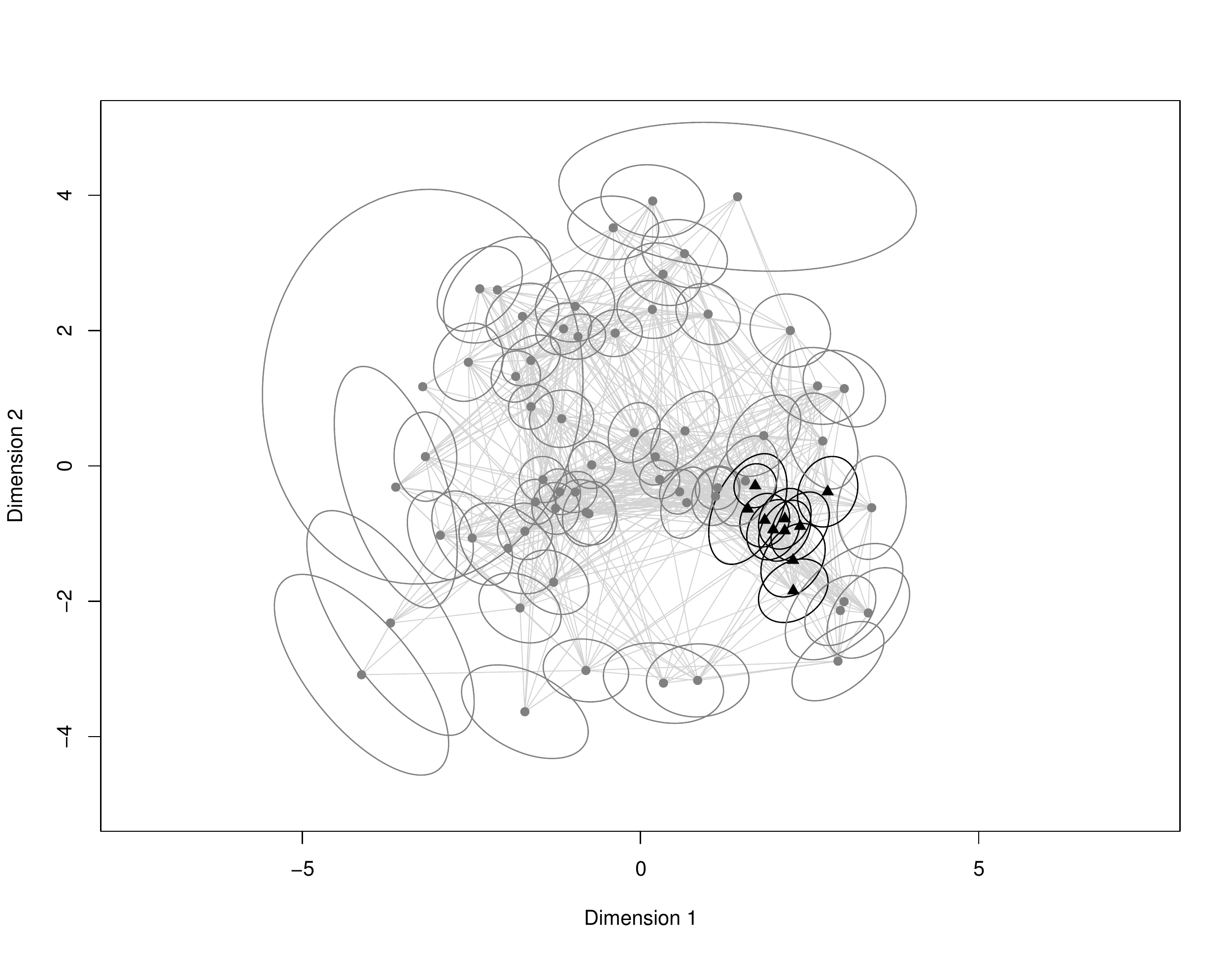}}&
\subfigure[Mixture of regressions latent position cluster model.]{\label{fig:box2}\includegraphics[width=60mm, height=65mm, angle=270]{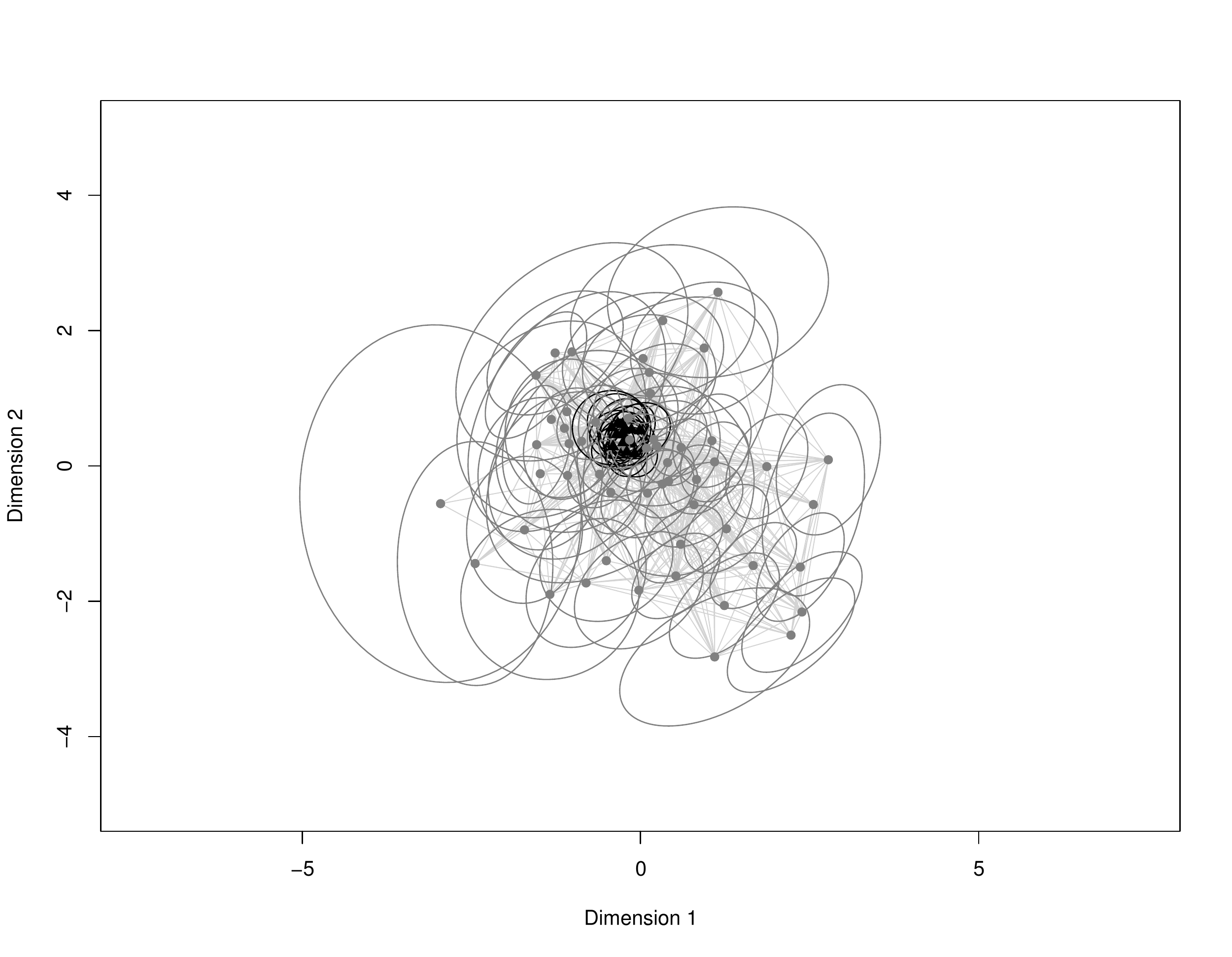}}\\
\subfigure[Simple latent position cluster model.]{\label{fig:box3}\includegraphics[width=60mm, height=65mm, angle=270]{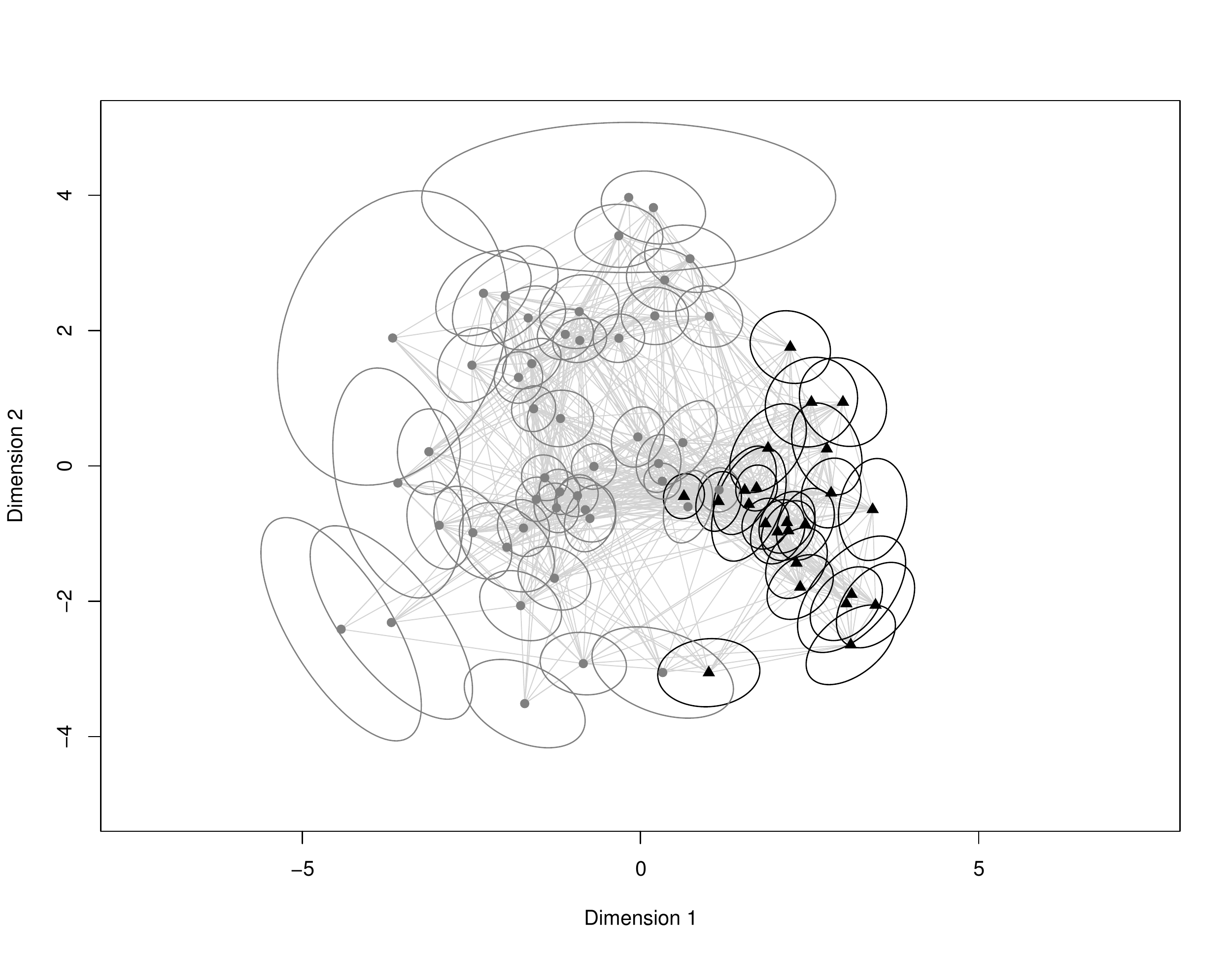}}&
\subfigure[Standard mixture of experts latent position cluster model.]{\label{fig:box4}\includegraphics[width=60mm, height=65mm, angle=270]{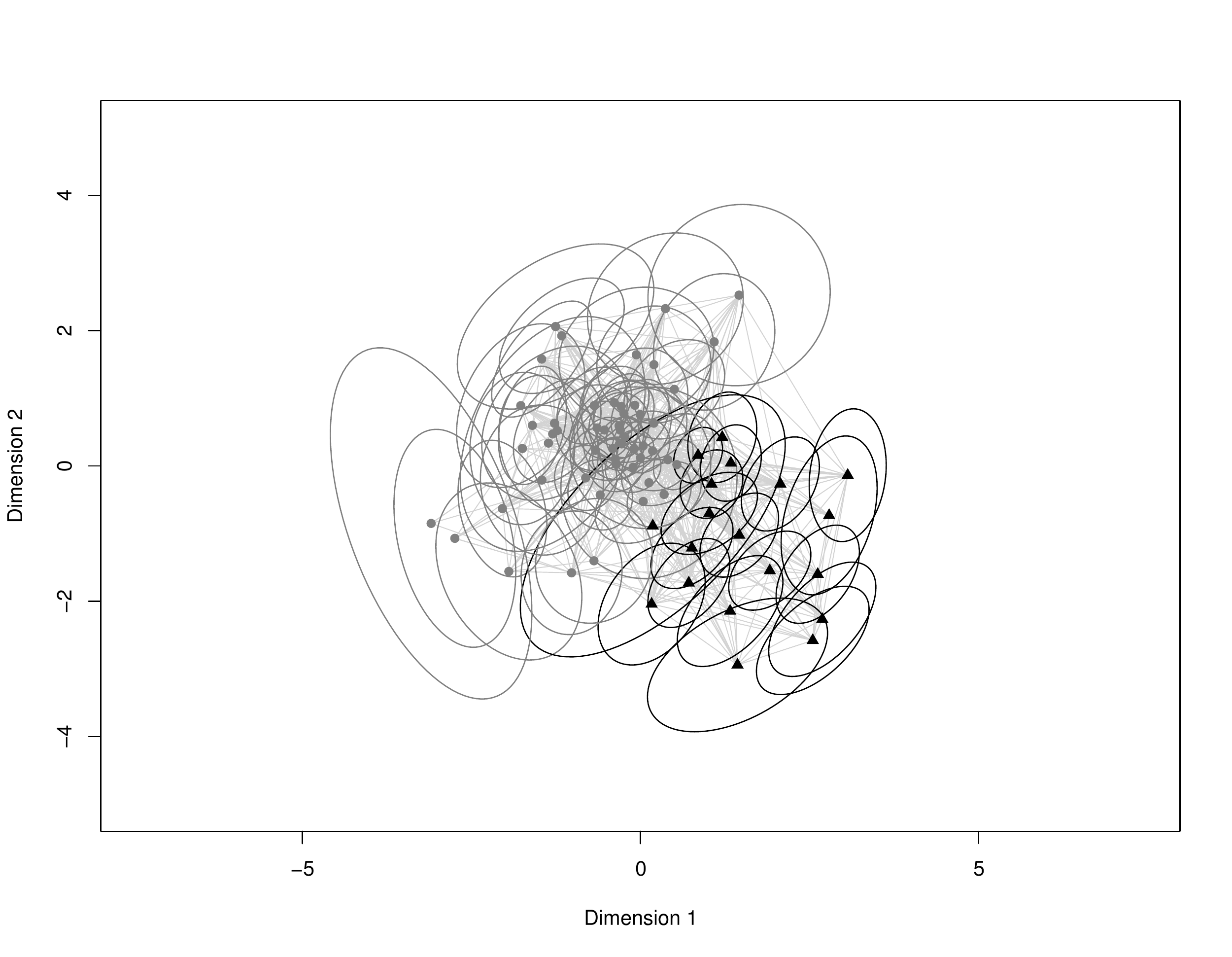}}
\end{tabular}
\end{center}
\vspace{-0.75cm}
\caption{Estimates of clusters and latent positions of the lawyers from the advice network data. The ellipses are 50$\%$ posterior sets illustrating the uncertainty in the latent locations. Lawyers who are members of the same cluster are illustrated using the same shade and symbol. Observed links between lawyers are also illustrated. }\label{fig:lawlocations}
\end{figure}

Under the ME model framework outlined in Section~\ref{sec:allMEmodels}, a suite of four mixtures of experts latent position cluster models is available. This suite of models was fitted to the advice network; data in this network detail links between lawyers who sought basic professional advice from each other over the previous twelve months. \cite{gormley10b} explore the coworkers network data set and the friendship network data set using similar methodology. Figure \ref{fig:lawlocations} illustrates the resulting latent space locations of the lawyers under each fitted model with $(G,\dold ) = (2,2)$.  These values were selected using BIC after fitting a range of latent position cluster models (with no covariates) to the network data only \citep{handcock07}. Table \ref{tab:lawcoeffs} details the resulting regression parameter estimates and their associated uncertainty for the four fitted models.

\begin{table}[t!]
\begin{center}
\caption{Posterior mean parameter estimates for the four mixtures of experts models fitted to the lawyers advice data as detailed in Figure \ref{fig:lawlocations}. Standard deviations are given in parentheses. Note that cluster 1 was used as the baseline cluster in the case of the cluster membership parameters. Baseline categories for the covariates are detailed in Table \ref{tab:lawcovs} \label{tab:lawcoeffs}}{}
\begin{tabular}{lcccc}
     & Model~(a) & Model~(b) & Model~(c) & Model~(d)
\\\hline 
\textbf{Link Probabilities}& \\\cline{1-1}
Intercept & 1.26 (0.10) & -2.87 (0.17)& 1.23 (0.10) & -2.65 (0.17)\\
Age        &               & -0.02 (0.004)&                    & -0.02 (0.004)\\
Gender   &               &  0.60 (0.09)&                    & 0.62 (0.09)\\
Office      &               & 2.02 (0.10)&                     & 1.97 (0.10)\\
Practice  &               & 1.63 (0.10)&                     & 1.57 (0.10)\\
Seniority &                & 0.89 (0.11)&                   & 0.81 (0.11)\\
Years     &                & -0.04 (0.005)&                     & -0.04 (0.005)\\
 & \\\hline 
\textbf{Cluster Memberships} & \\\cline{1-1}
Intercept    & -1.05 (1.75)& 0.94 (0.79) & -0.62 (1.23) & 1.27 (1.29)\\
Age            &                &                    & -0.09 (0.04) & -0.14 (0.06)\\
Office (=1) &                &                   & 1.94 (1.02)  & 2.40 (1.14)\\
Office (=2) &                &                   & -2.08 (1.09) & -0.97 (1.19)\\
Practice     &                &                   & 3.18 (0.85)  & 2.14 (1.08)\\
 & \\\hline 
\textbf{Latent Space Model} & \\\cline{1-1}
Cluster 1 mean     & -0.50 (0.52) & 0.09 (0.19) & -1.09 (0.31) & -0.54 (0.21)\\
                             &  0.21 (0.58) & -0.09 (0.26) & 0.40 (0.28)    & 0.40 (0.20)\\
Cluster 1 variance &  3.35 (1.29) & 2.12 (0.77) & 3.19 (0.58) & 1.25 (0.34)\\
 & \\
Cluster 2 mean     & 1.66 (0.92) & -0.24 (0.20) & 2.10 (0.30) & 1.32 (0.51) \\
                             & -0.67 (0.58) & 0.35 (0.23) & -0.77 (0.30)    & -0.98 (0.47)\\
Cluster 2 variance & 1.29  (1.58) & 0.27 (0.68) & 1.16 (0.40) & 1.63 (0.69)\\
 & \\\hline 
AICM & -3644.24& -3346.87 & -3682.71 & -3325.95\\ 
\end{tabular}
\end{center}
\end{table}

The models are compared through the AICM, the posterior simulation-based analogue of Akaike's Information Criterion (AIC) \citep{akaike:1973, raftery:etal:2008}. In this implementation the optimal model is that with the highest \index{AICM} AICM and is the model with covariates in the link probabilities and in the component weights.  The results of the analysis show some interesting patterns. The coefficients of the covariates in the link probabilities are very similar in the models (b) and (d) in Table~\ref{tab:lawcoeffs}. These coefficients indicate that a number of factors have a positive or negative effect on whether a lawyer asks another for advice. In summary, lawyers who are similar in seniority, gender, office location and practice type are more likely to ask each other for advice. The effects of years and age seem to have a negative effect, but these variables are correlated with seniority and with each other, so their marginal effects are more difficult to interpret.

Importantly, the latent positions are very similar in models (a) and (c) which do not have covariates in the link probabilities and models (b) and (d) which do have covariates in the link probabilities. This can be explained because of the different role that the latent space plays in the models with covariates in the link probabilities and those that do not have such covariates. When the covariates are in the link probabilities, the latent space is modelling the network structure that could not be explained by the link covariates, whereas in the other case the latent space is modelling much of the network structure.

Interestingly, in the model with the highest AICM value, there are covariates in the cluster membership probabilities as well as in the link probabilities. This means that the structure in the latent space, which is modelling what could not be explained directly in the link probabilities, has structure that can be further explained using the covariates. The office location, practice and age of the lawyers retain explanatory power in explaining the clustering found in the latent social space.

The difference in the cluster membership coefficients in models (c) and (d) is due to the different interpretation of the latent space in these models. However, it is interesting to note that in this application the signs of the coefficients are identical because the cluster memberships shown for these models in Figure~\ref{fig:box3} and Figure~\ref{fig:box4} are similar; this phenomenon does not hold generally \citep[see][Section 5.3]{gormley10b}.

The results of this analysis offer a cautionary message in automatically selecting the type of mixture of experts latent position cluster model for analyzing the lawyer advice network. The role of the latent space in the model is very different depending on how the covariates enter the model. So, if the latent space is to be interpreted as a social space that explains network structure, then the covariates should not directly enter the link probabilities. However, if the latent space is being used to find interesting or anomalous structure in the network that cannot be explained by the covariates, then one should consider allowing the covariates enter the cluster membership probabilities.

\subsection{Software} \label{sec:software}

As demonstrated in this section, the approach to fitting an ME model depends on the application setting and on the form of the ME model itself.  Therefore, a single software capable of fitting any ME model is not currently available.

In R \citep{R:2018}, the \texttt{MEclustnet}\index{R package!MEclustnet@{\sf MEclustnet}} package \citep{gormley18} fits the mixture of experts latent position cluster model detailed in Section~\ref{sec:networkeg}. The \texttt{flexmix}\index{R package!flexmix@{\sf flexmix}} package \citep{gruen:leisch:2008} has model fitting capabilities for a range of mixture of regression models, which include covariates (or concomitant variables), as does the \texttt{mixreg} package \citep{mixreg14}.\index{R package!mixreg@{\sf mixreg}} Additionally, \texttt{mixtools}\index{R package!mixtools@{\sf mixtools}} \citep{benaglia09} facilitates fitting of a $G = 2$ mixture of regressions model in which the component weights are modelled as an inverse logit function of the covariates. The cluster weighted models which are closely related to ME models can be fitted using the \texttt{flexCWM} package \citep{mazza17}.\index{R package!flexCWM@{\sf flexCWM}} All packages are freely available through the Comprehensive R Archive Network (CRAN) at {\tt https://cran.r-project.org}.

In MATLAB, the  \texttt{bayesf}\index{MATLAB package!bayesf@{\sf bayesf}}  package \citep{fru:bayesf} allows to estimate   a broad
range  of  mixture models using either finite mixtures, mixtures of experts or Markov switching models as a model for the hidden group indicators $\zm$.

In terms of other softwares, the \texttt{FMM} procedure in SAS also facilitates ME model fitting, and stand alone softwares such as \texttt{Latent GOLD} \citep{vermunt:2005} and \texttt{Mplus} \citep{muthen:2011} fit closely related latent class models.

\section{Identifiability of Mixtures of Experts Models} \label{sec:identifiability}


 For a finite mixture distribution one has to distinguish
 three types of non-identifiability\index{identifiability} \citep[Section~1.3]{fru:book}:  invariance to relabelling   the
components of the mixture distribution (the so-called label switching problem),
 non-identifiability due to potential overfitting 
  and  generic non-identifiability which  occurs
only for certain classes of mixture distributions.

 Consider a standard mixture distribution with $G$   components with non-zero weights
 $\eta_1, \ldots, \eta_G$  generated by distinct parameters $ \theta_1, \ldots,\theta_G$.
 Assume that  for all possible realisations $y$ from this mixture distribution 
  the  identity
\begin{eqnarray*}
\sum_{g=1}^G \eta_g f_g( y| \theta_g) = \sum_{g=1}^{G^\star} \idestar{\eta}_g f _g( y| \idestar{\theta}_g)
\end{eqnarray*}
holds   where the right-hand  side is a  mixture distribution from the same family with  $G^\star$   components with non-zero weights
 $\idestar{\eta}_1, \ldots, \idestar{\eta}_{G^\star}$  generated by distinct parameters $ \idestar{\theta}_1, \ldots,\idestar{\theta}_{G^\star}$.
 Then generic identifiability\index{identifiability} implies that  ${G^\star}=G$
 and the two mixtures'  parameters  $\theta=(\eta_1, \ldots,\eta_G, \theta_1, \ldots,\theta_G)$ and  $\theta^\star=(\idestar{\eta}_1, \ldots,\idestar{\eta}_G, \idestar{\theta}_1, \ldots,\idestar{\theta}_G)$ are identical
up to relabelling the component indices.
Common finite mixture distributions such as  Gaussian  and  Poisson mixtures are generically identified, see   \citet{tei:ide63},
 \citet{yak-spr:ide},  and \citet{cha:mix} for a detailed discussion. 

 Discrete mixtures   often suffer from generic non-identifiability for certain parameter configurations, well-known examples being mixtures of binomial distributions (see Section~\ref{sec13:ibin})  and  mixtures of multinomial distributions
 \citep{gru-lei:ide}. Somewhat unexpectedly, mixtures  of regression models  suffer from  generic non-identifiability \citep{hen:ide,gru-lei:fin},\index{identifiability}
  as will be discussed in more detail in Section~\ref{sec13:ide}. Little is known about  generic identifiability of mixtures of experts models and some results are presented in Section~\ref{sec_ideME}. However, ensuring generic identifiability for general ME models  remains a  challenging issue.

Identifiability problems for mixture with nonparametric components are discussed in Chapter~14 of this volume.

\subsection{Identifiability for mixtures of binomials}  \label{sec13:ibin}

For binomial mixtures  the component densities arise from $\Bino(\BinoT,\pl)$-distributions, where $\BinoT$ is commonly
assumed to be known, whereas $\pl$ is heterogeneous across the components:\index{identifiability}
\begin{eqnarray}
\rvY  \sim \eta_1 \Bino(\BinoT,\plt{1}) + \cdots + \eta_G \Bino(\BinoT,\plt{G}). \label{mix:fbinn:eq1}
\end{eqnarray}
 The probability mass function (pmf) of this mixture takes on $\BinoT+1$ different support points:
\begin{eqnarray}
p(\ydens|\thmod)  = \Prob ( \rvY= \ydens|\thmod)
 = \sum_{g=1}^G \eta_g \Bincoef{\BinoT}{\ydens}
\plt{g}^\ydens (1-\plt{g})^{\BinoT-\ydens} ,  \quad  \ydens=0, 1, \ldots,  \BinoT, \label{mix:binoinP}
\end{eqnarray}
with  $2G-1$ independent parameters  $\thmod=(\plt{1},\ldots,\plt{G},\eta_1,\ldots,\eta_G)$,  with  $\eta_G=1-\sum_{g=1}^{G-1} \eta_g$.

Given data $\ym=(y_1,\ldots, y_n)$ from mixture (\ref{mix:fbinn:eq1}), the only information available to estimate $\thmod$ are  $ \BinoT$ (among the
 $ \BinoT+1 $ observed) relative frequencies  $\fnarg{n}{ \rvY= \ydens}$ ($\ydens=0, 1, \ldots,  \BinoT$).
  As $n \rightarrow \infty$ (while $\BinoT$ is fixed),
$\fnarg{n}{ \rvY= \ydens}$ converges to  $\Prob ( \rvY= \ydens|\thmod) $ by the law of large numbers, but the number of  support points  remains fixed. Hence, the data provide only  $ \BinoT$  statistics, given by the  relative frequencies,  to estimate  $2G-1$  parameters. Simple counting yields the following necessary condition for identifiability for a    binomial mixture, which has been shown by \citet{tei:ide61} to be also sufficient:
\begin{eqnarray} \label{idemixbin}
 2G-1  \leq  \BinoT  \qquad \Leftrightarrow  \qquad  G  \leq ( \BinoT + 1)/2.
\end{eqnarray}
 Consider, for  illustration, a mixture of  two binomial distributions,
\begin{eqnarray}  \label{Bino2}
\rvY \sim \eta \times \Bino(\BinoT,\plt{1})+ (1-\eta) \times \Bino(\BinoT,\plt{2}),
\end{eqnarray}
with three unknown parameters  $\thmod =( \eta,\plt{1},\plt{2})$ 
 and assume that the population indeed contains two different groups, i.e. $\plt{1} \neq \plt{2}$ and  $\eta>0$.  Assuming  $\BinoT=2$
  obviously violates   condition  (\ref{idemixbin}).  Lack of identification  can be verified directly from  the pmf
which is different from zero only for the three outcomes $y\in \{ 0,1,2 \}$:
\begin{eqnarray}  \label{equprob}
 && \Prob(\rvY=0|\thmod) = \eta (1-\plt{1})^2 +   (1- \eta) (1-\plt{2})^2,\\
  && \Prob(\rvY=1|\thmod) = 2 \eta \plt{1} (1-\plt{1}) +  2 (1- \eta) \plt{2}(1-\plt{2}), \nonumber \\
 && \Prob(\rvY =2|\thmod) = \eta \plt{1}^2 +   (1- \eta)  \plt{2}^2. \nonumber
\end{eqnarray}
Since $\sum_y \Prob(\rvY=y|\thmod)=1$,  only two linearly independent equations remain to identify the three parameters  $( \eta,\plt{1},\plt{2})$.
Hence  parameters  $\thmod=(\plt{1},\plt{2},\eta) \neq \idestar{\thmod} =(\idestar{\plt{1}},\idestar{\plt{2}},\idestar{\eta})$
 fulfilling equations (\ref{equprob}) exist  which imply the same distribution for $\rvY$, i.e.:
 $ \Prob (\rvY= \ydens|\thmod) =   \Prob ( \rvY= \ydens|\idestar{\thmod })$, $\forall  y=0,1,2$,
but are not related to each other by simple relabelling of the component indices.

Such generic non-identifiability\index{identifiability} severely  impacts statistical estimation of  the mixture parameters $\thmod$
 from  observations  $\ym=(y_1,\ldots, y_n)$, even if $G$ is known,  and goes far beyond label switching.
Assume, for illustration,  that $\ym$ is the  realisation of a random sample  $(Y_1,\ldots, Y_n)$
 from the two-component binomial mixture (\ref{Bino2})  with  $\BinoT=2$   and
 true parameter $\thmod \true=(\plt{1}\true ,\plt{2}\true ,\eta \true) $
 and consider the corresponding  observed-data likelihood
 $ p(\ym|\thmod) =\prod_{i=1}^n  \Prob(\rvY_i=y_i|\thmod)  . $
 Generic non-identifiability  of the underlying mixture distribution implies that  the observed-data
 likelihood is the same  for  any pair $ \thmod \neq \idestar{\thmod}$ of distinct parameters  satisfying (\ref{equprob}), for any possible  sample $\ym$ in the sampling space $\samspace = \{ 0,1,2\}^n$,~i.e.:
 $ p(\ym|\thmod)  =    p(\ym| \idestar{\thmod}) ,  \forall  \ym \in \samspace .$
 Since this holds for  arbitrary sample size $n=1, 2, \ldots$,  the true  parameter $\thmod \true$ cannot be recovered,
 even if   $n \rightarrow \infty$, and  both maximum likelihood estimation as well Bayesian inference suffer from
 non-identifiability problems for such a mixture.

This example motivates the following more formal definition of generic  non-identifiability.\index{identifiability}
For a given $\thmod$,  any subset $\ideU (\thmod)$ of the parameter space $\Theta$ of a mixture model, defined as
$\ideU (\thmod) = \left\{ \idestar{\thmod} \in \Theta: p(\ym |\idestar{\thmod})= p(\ym |\thmod),   \forall \ym \in \samspace  \right\},$
is called a non-identifiability set, if  it  contains at least one point  $\idestar{\thmod}$ which is not related to $\thmod$ by simple relabelling of  the component indices. 
Let $\thmod \true $  be the true  parameter  value  of a mixture model with $G$  distinct  parameters (i.e. $\theta_g\neq   \theta_{g'}$, for  $g\neq  g'$).  If $\ideU (\thmod \true)$  is a non-identifiability set in the sense defined above, then
  $\thmod \true $   cannot be recovered  from  data, even as   $n$  goes to infinity.

\begin{figure}[t!]
\begin{center}
\begin{tabular}{cc}
\scalebox{0.3}{\includegraphics{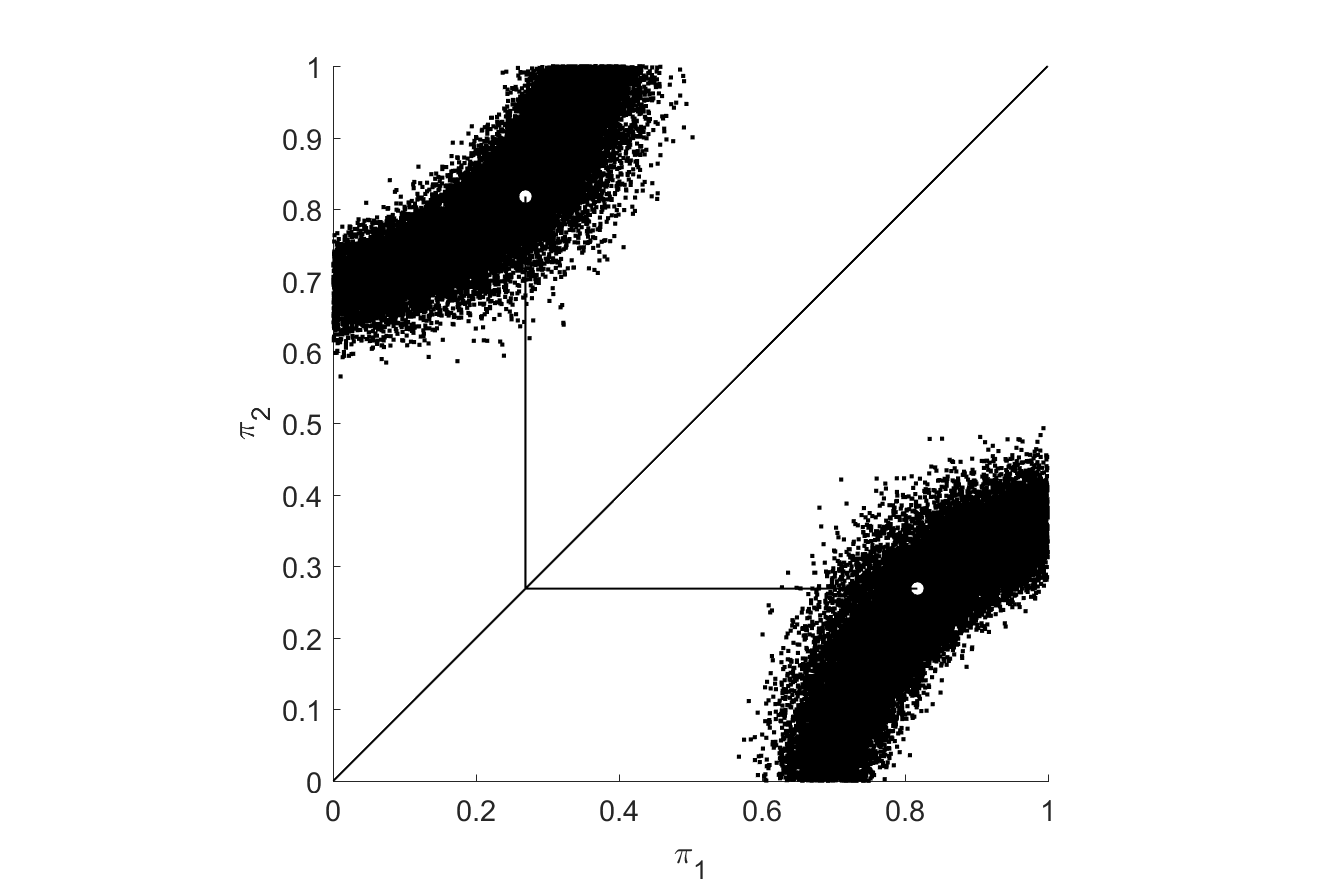}}& \scalebox{0.3}{\includegraphics{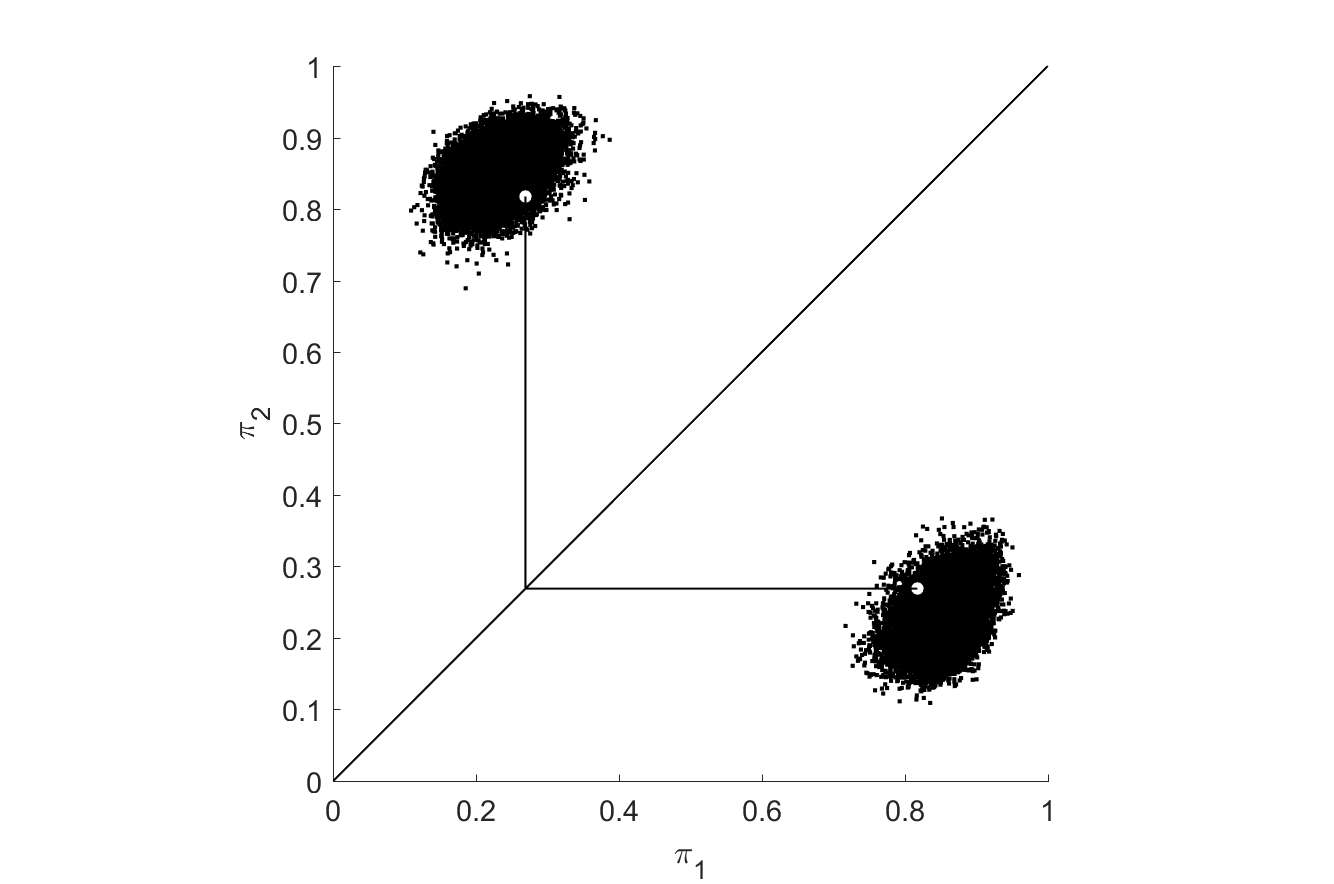}}\\
\scalebox{0.3}{\includegraphics{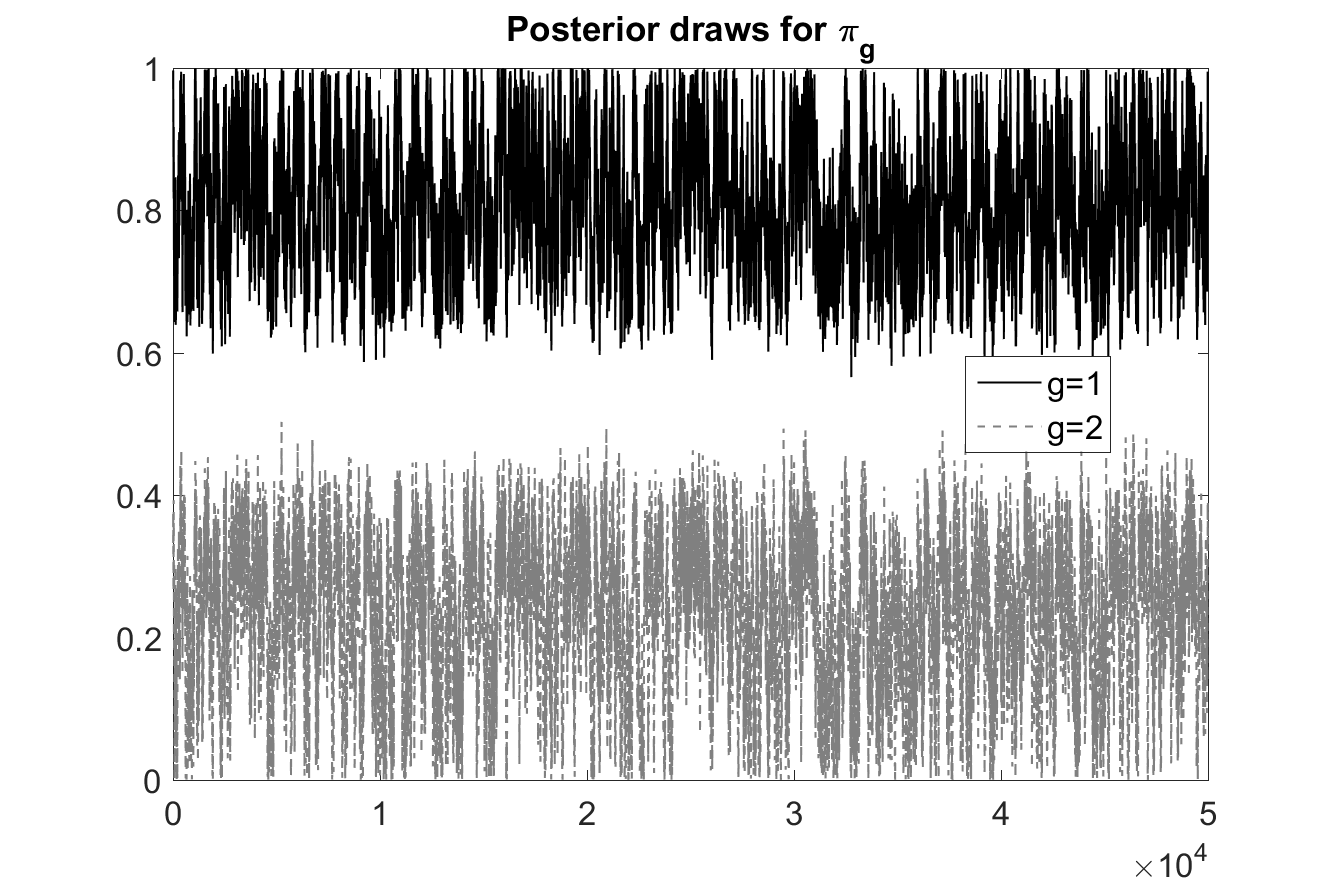}}& \scalebox{0.3}{\includegraphics{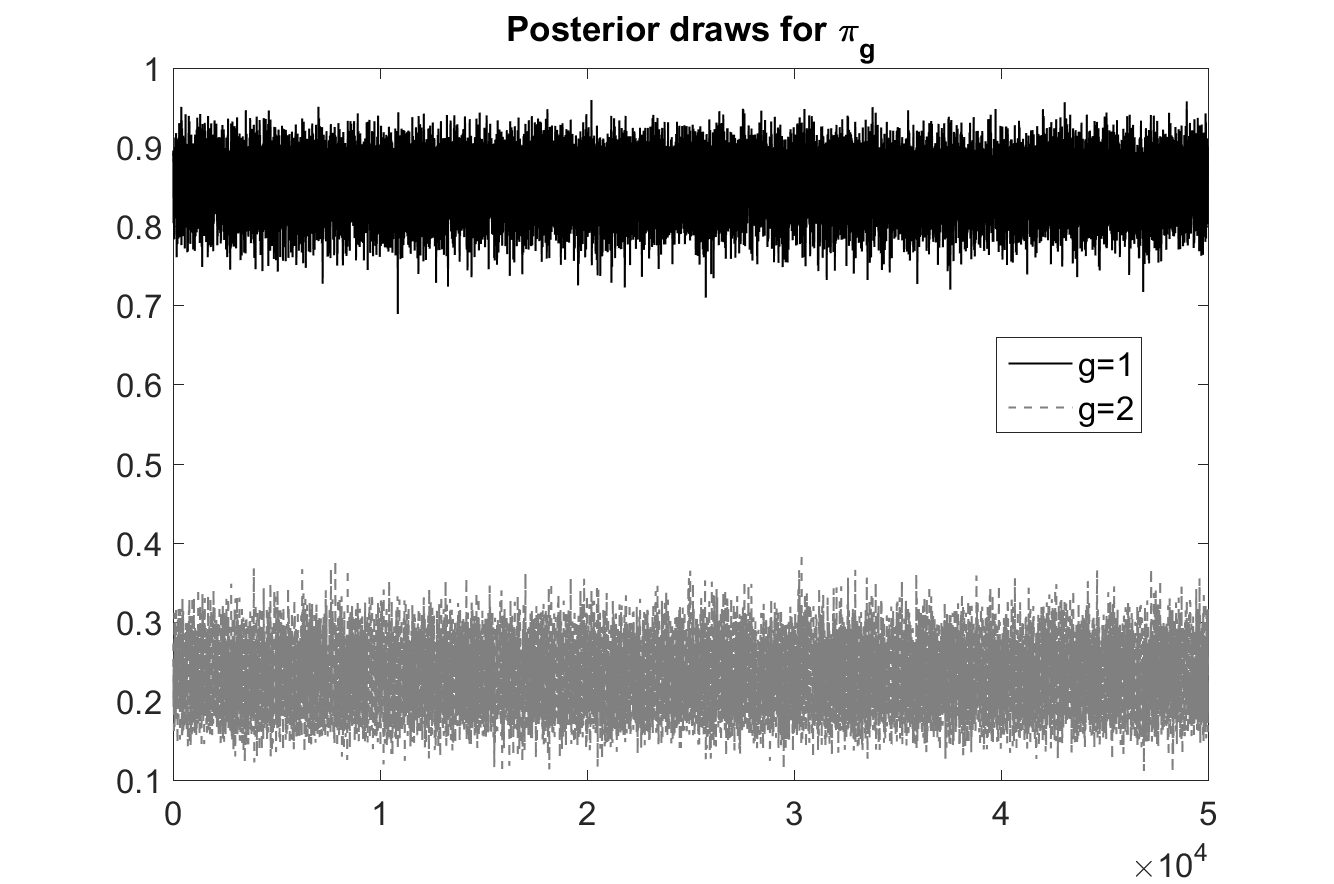}}
\end{tabular}
\end{center}
\vspace{-0.75cm}
\caption{MCMC inference for data simulated from a mixture of two binomial distributions with  $\BinoT=2$ (left-hand side)
  and  $\BinoT=5$    (right-hand side).  Top: scatter plot of  $\pi_{1}$ versus $\pi_{2}$ (true values indicated by a circle). Bottom:
posterior draws of  the group-specific probabilities  $\pi_{1}$ and $\pi_{2}$ after resolving label switching  in
the scatter plot of $\pi_{1}$ versus $\pi_{2}$ through $k$-means clustering.}\label{bino:mcmc}
\end{figure}

  Such generic non-identifiability  has important   implications for practical mixture analysis. 
  For finite $n$,   the observed-data  likelihood  function  $p(\ym|\thmod)$ has a ridge close to $\ideU (\thmod \true)$  instead of  $G!$  isolated modes and no unique maximum, leading to inconsistent estimates of   $\thmod \true$.
  In a Bayesian framework, this leads to a posterior distribution that does not concentrate around  $G!$ isolated, equivalent modes  as $n$ increases,
 as  for identifiable\index{identifiability} models (see  Chapter~4, Section~4.3). Rather, the posterior concentrates over the entire non-identifiability set
  $\ideU (\thmod \true)$ which has a  complex geometry and  can be represented as the union of $G!$   symmetric subspaces, see e.g.  Figure~\ref{bino:mcmc} for a binomial mixture with $G=2$ and $\BinoT=2$.
 The  prior $p(\thmod)$ provides information  beyond the data  and might  influence how  the posterior   concentrates  on
  each of these $G!$ subspaces $\ideU (\thmod \true)$, in particular, if  the prior $p(\theta)$ is not constant  over 
 $\ideU (\thmod )$.

While generic non-identifiability  has important  practical implications  for mixture analysis,
 it  is  rarely as easily diagnosed as for mixtures of binomial distributions
and can easily  go unnoticed for more complex mixture models,  in particular  for  maximum likelihood  estimation,  whereas MCMC based Bayesian inference often provides indications of  potential identifiability  problems, as the following example demonstrates.\index{identifiability}

 \paragraph*{MCMC inference for an example: a mixture of binomial distributions}
 $\:$\\
 $\:$\\
 For further illustration, we perform MCMC inference
 (based on 10,000 draws after a burn-in of 5,000 iterations) for two data sets simulated from a mixture of two binomial distributions with
  $\logit \pi_1=-1$ and  $\logit \pi_2=1.5$ using random permutation sampling as explained in Chapter~5, Section~5.2.
    We assume that $G=2$ and $\BinoT$ is known,  whereas  all other parameters   in  mixture (\ref{Bino2}) are unknown.  Bayesian inference is based on the following priors:  $\pi_{g} \sim \Uniform(0,1)$, and $(\eta_1,\eta_2) \sim \Dir (1,1)$.    The two data sets were generated with, respectively,   $\BinoT=2, n=250$  and   $\BinoT=5, n=100$, implying the same  total number   $n \times \BinoT =500$ of experiments.

 MCMC inference is summarised in Figure~\ref{bino:mcmc}, showing  scatter plot of  $\pi_{1}$ versus $\pi_{2}$ 
 for both values of
 $\BinoT$.    For   $\BinoT=5$, the mixture is generically identified and the posterior draws concentrate  around  two symmetric  modes, centered  at the
 true values 
 $(0.269,0.818)$ and $(0.818,0.269)$. Non-identifiability due to  label switching is resolved by applying $k$-means clustering to the posterior draws,  see the lower part of Figure~\ref{bino:mcmc}   showing  identified posterior draws of  the group-specific success probabilities $\pi_{1}$ and  $\pi_{2}$.

 For   $\BinoT=2$,  a similar  scatter plot of  $\pi_{1}$ versus $\pi_{2}$
clearly indicates severe identifiability issues, showing that the posterior  draws arise from two symmetric unidentifiability sets, rather than  concentrating around two symmetric modes centered at the true values.  When we apply $k$-means clustering to resolve label switching, we obtain the posterior draws of  the success probabilities $\pi_{1}$  and  $\pi_{2}$  shown in the lower part of Figure~\ref{bino:mcmc},  also indicating problems  with identifying  $\pi_{1}$  and  $\pi_{2}$ from the data for $\BinoT=2$.

\subsection{Identifiability for mixtures of regression models} \label{sec13:ide}

\begin{figure}[t!]
\begin{center}
\begin{tabular}{cc}
\scalebox{0.3}{\includegraphics{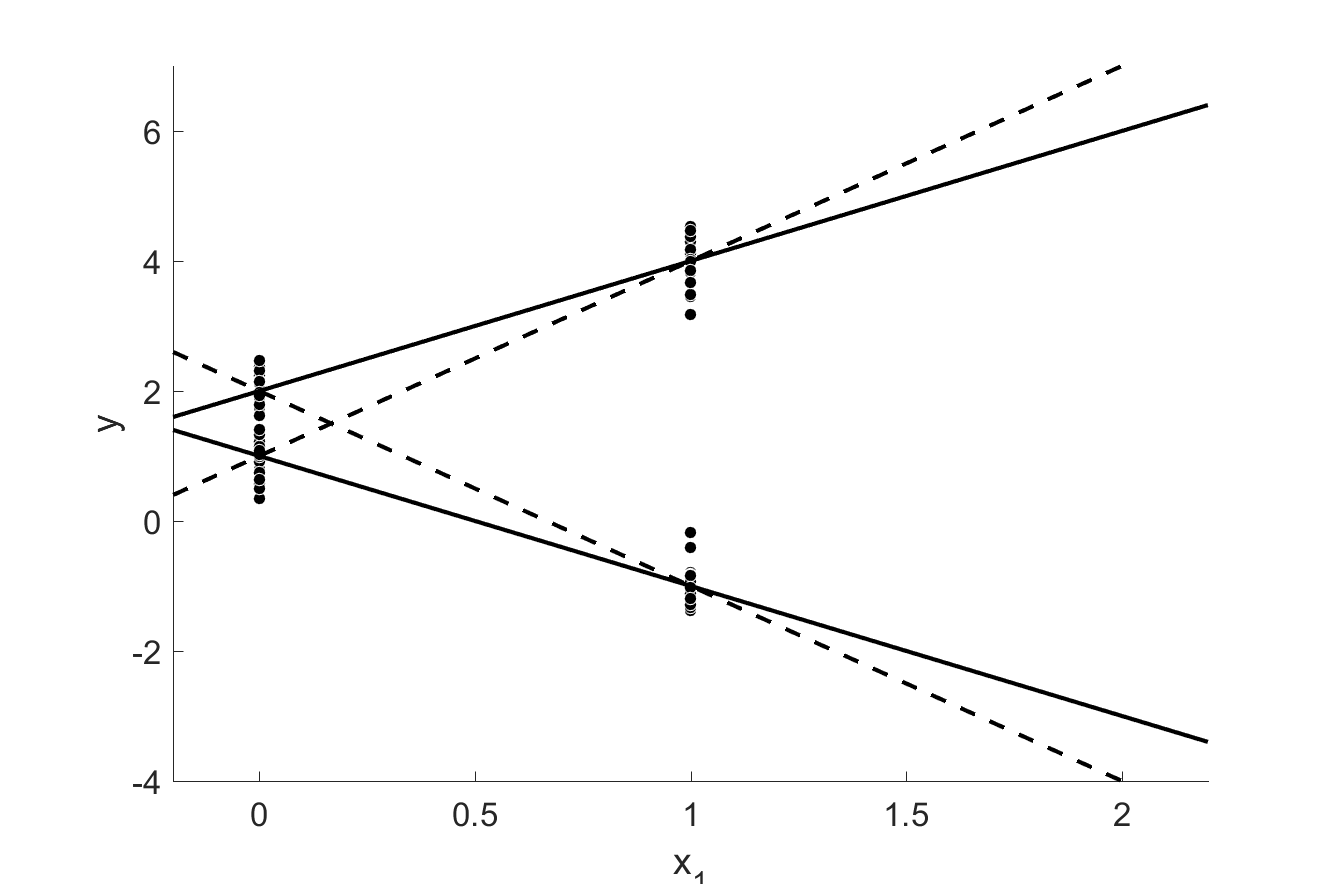}} & \scalebox{0.3}{\includegraphics{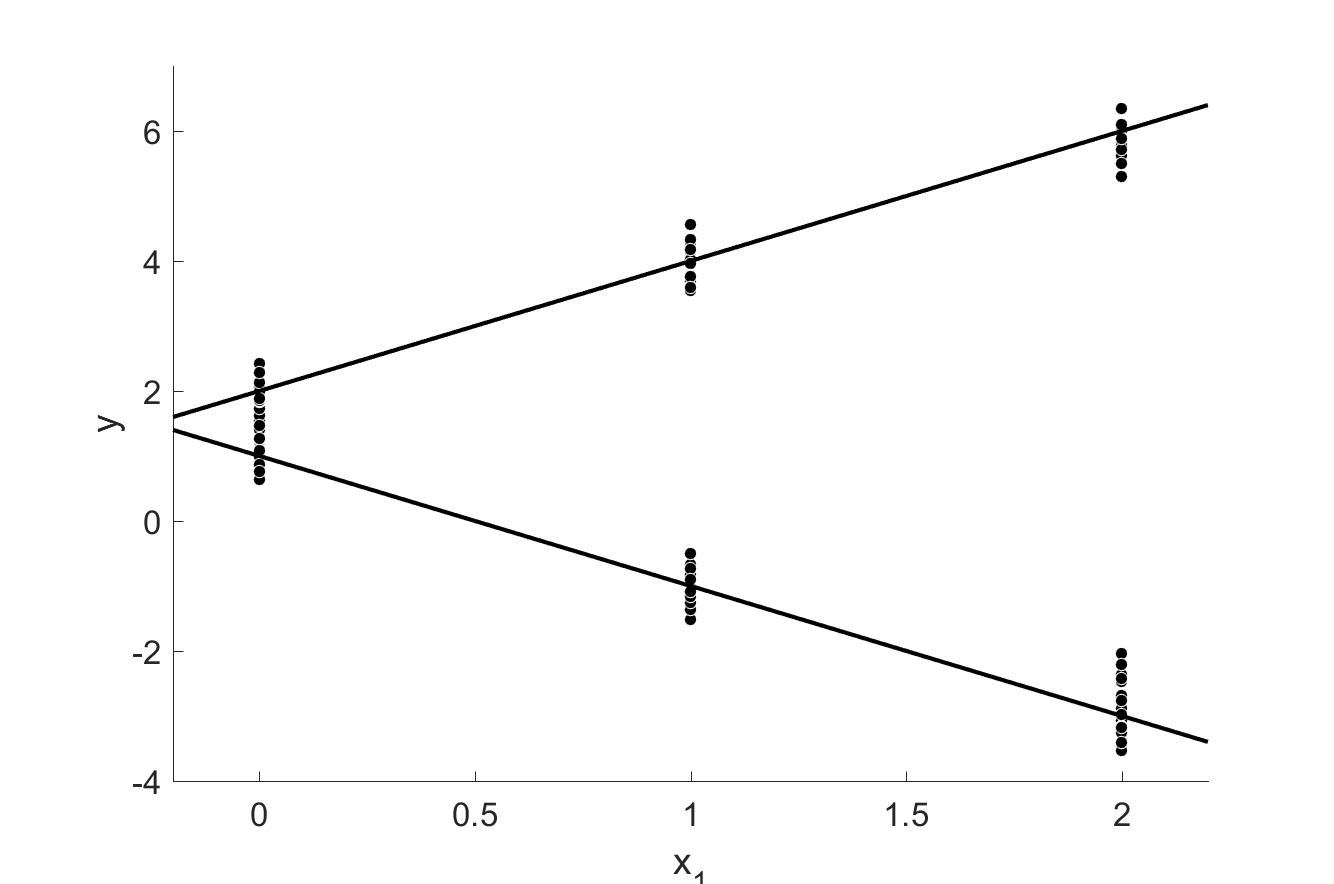}}
\end{tabular}
\end{center}
\vspace{-0.75cm}
\caption{Data simulated from a mixture of two regression lines under   \textit{Design~1} (left-hand side) and
 \textit{Design~2} (right-hand side).   The full lines indicate the true underlying model used to generate 100 data points (black dots).  For  the unidentified \textit{Design~1}, a second solution exists which is indicated by the dashed lines.}\label{regmix:data}
\end{figure}

 Consider a mixture of  $G$ regression models for $i=1, \ldots, n$
outcomes $y_i$, arising from $G$  different groups,
\begin{eqnarray} \label{mixreg1}
y_i  | \Xbetatilde_i  \sim \sum_{g=1}^G \eta_g \phi ( y|\muswt{i}{g} (\Xbetatilde_i),\sigmaerrsw{g})
\end{eqnarray}
where for each  $g=1,\ldots, G$,  the group-specific mean  $\muswt{i}{g} (\Xbetatilde_i)  = \Xbetatilde_i  \betavsw{g}$ depends  on
 a group-specific  regression parameter  $\betavsw{g}$   and  on the $(1 \times (q+1))$-dimensional row vector $ \Xbetatilde_i$   containing the $q$ covariates  $x_i$ and a constant. 
For a fixed design point $ \Xbeta= \Xbetatilde_i $, (\ref{mixreg1}) is a standard finite Gaussian mixture distribution and as such generically identified.\index{identifiability}
Hence, if the identity
\begin{eqnarray}
\sum_{g=1}^G \eta_g \phi (y| \muswt{i}{g} (\Xbeta) ,\sigmaerrsw{g}) =
\sum_{g=1}^G \idestar{\eta}_g \phi ( y| \idestar{\muswt{i}{g}} (\Xbeta) ,\idestararg{\sigmaerrsqrtsw{g}}{2}) , \label{eqsreg}
\end{eqnarray}
 holds, then  the two mixtures  are related to each other by relabelling, i.e.
 $\idestar{\muswt{i}{g}} (\Xbeta) =\muswt{i}{\perm_x(g)} (\Xbeta)  = \Xbeta  \betavsw{\perm_x(g)}$,
$\idestararg{\sigmaerrsqrtsw{g}}{2}=\sigmaerrsw{\perm_x(g)}$,  and $ \idestar{\eta}_g=\eta_{\perm_x(g)}$ for $g=1,\ldots, G$,  for some
permutation $\perm_x \in \mathfrak{S}(G)$, where  $\mathfrak{S}(G)$ denotes  the set of the $G!$ permutations of $\{1,\ldots,G\}$.
Note that $\perm_x$ depends on  the covariate $\Xbeta$  and that there is no guarantee that  $\perm_x$ is identical
across different values of  $\Xbeta$ which can cause {\em intra-component label switching}. One such example is displayed
on the left-hand side of Figure~\ref{regmix:data}.

Nevertheless,  assume for the moment  that   $\perm_{x} \equiv \perm_{\star}  $ is the same  for all possible covariates $\Xbeta$. Then (\ref{eqsreg})
implies  $ \Xbetatilde_i \idestar{\betavsw{g}} =
 \Xbetatilde_i  \betavsw{\perm_{\star}(g)}$ for all
  $i= 1,\ldots,n$ and
  $\Xbetamat \idestar{\betavsw{g}} = \Xbetamat  \betavsw{\perm_{\star}(g)}$
 where the rows of the  matrix $ \Xbetamat$ are equal to $\Xbetatilde_{1}, \ldots, \Xbetatilde_{n }$. If  the usual condition in regression modelling
is satisfied that  $\Xbetamat ^\top \Xbetamat$ has full rank, then it follows immediately that the regression coefficients are
determined up to relabelling:
$  \idestar{\betavsw{g}}  =  \betavsw{\perm_{\star}(g)}$.

Hence, generic identifiability for a mixture of regressions model can be verified through sufficient conditions  guaranteeing  that $\perm_x$ is  indeed identical\index{identifiability}
across  all values of   $\Xbeta$.  Mathematically,   one such condition is the assumption that
 either   the error variances $\sigmaerrsw{1}, \ldots,  \sigmaerrsw{G}$ or the weights  $\eta_{1}, \ldots, \eta_{G}$ satisfy a strict order constraint.  However, in practice such constraints are rarely fulfilled and forcing an order constraint on one coefficient  does not necessarily prevent label switching for the other coefficients in Bayesian posterior sampling,  see e.g. \citet[Section~2.4]{fru:book}.

Hence, several papers focused on conditions for generic identifiability through the regression part of the model \citep{hen:ide,gru-lei:ide,gru-lei:fin}.
Assume that  the covariates $\Xbetatilde_i$  take $\ndes$ different values  in a design space $\{ \Xbeta_1, \ldots, \Xbeta_\ndes \}$ for  the  observed outcome $y_i$, for $ i=1,\ldots,n$. Identifiability through the  regression part requires  enough variability in the design space
and is guaranteed under  so-called {\em coverage conditions}. These conditions require that the number of clusters $G$ is exceeded by the minimum number of distinct  $q$-dimensional hyperplanes  needed to cover the covariates (excluding the constant).
For $q=1$, for instance, the  coverage condition is satisfied, if the number of design points $p$ (i.e. the number of distinct values of the univariate covariate) is larger than
the number of clusters $G$.
These identifiability conditions go far beyond the usual condition that $\Xbetamat ^\top \Xbetamat $ has full rank and  are often violated for regression models with too few design points, a common example  being regression models with 0/1 dummy variables as covariates which are identifiable for $G=1$, but not for $G>1$, as the following examples with $q=1$ demonstrate.

For illustration, we consider the following special case of the mixture of regressions model  (\ref{mixreg1})  investigated  in \citet[Section~3.1]{gru-lei:fin}:
  \begin{eqnarray} \label{mixreg2}
y_i  \sim  0.5  \phi (y| \muswt{i}{1}(\Xbetatilde_i) ,0.1) + 0.5  \phi (y|  \muswt{i}{2} (\Xbetatilde_i),0.1) ,
\end{eqnarray}
with  covariate vector  $ \Xbetatilde_i=(1 \,\, d_i)$   and  group-specific regression parameters $ \betavsw{1}=(  2 \,\,\,  2 ) ^\top$ and  $\betavsw{2}=(   1 \,\,  -2 ) ^\top$.
 We consider two different regression designs, \textit{Design~1}  where $d_i$ is a 0/1 dummy variable capturing   the effect of gender (with female as baseline) and \textit{Design~2}  where $d_i$ captures a time effect over 3 periods (with $t=0$ serving  as baseline):
\begin{eqnarray*}
&& \mbox{\textit{Design~1}: }  \Xbeta_1=\left(\begin{array}{cc}1 & 0 \end{array}\right), \,\, \Xbeta_2=\left(\begin{array}{cc}1 &1 \end{array}\right), \\
&& \mbox{\textit{Design~2}: }  \Xbeta_1=\left(\begin{array}{cc}1 & 0 \end{array}\right), \,\, \Xbeta_2=\left(\begin{array}{cc}1 & 1 \end{array}\right), \,\,
\Xbeta_3=\left(\begin{array}{cc}1 & 2 \end{array}\right).
\end{eqnarray*}
In the following,  it is verified that  mixture (\ref{mixreg2}) is generically identified  under  \textit{Design~2}  (which contains three design points), but generically unidentified under  \textit{Design~1}  (which contains only two design points).

We first consider \textit{Design~1}.   According to  (\ref{eqsreg}),
  $\muswt{j}{1}$ and $\muswt{j}{2}$ are identified for $j=1$ and $j=2$ up to label switching
  arising from  two permutations  $\perm_1$ and $\perm_2$, where we may assume without loss of generality that  $\perm_1$ is equal to the identity:
  \begin{align} \label{regperm1}
&  \Xbeta_1  \betavsw{1} = \muswt{1}{1}, &     &\Xbeta_1  \betavsw{2} = \muswt{1}{2} ,   \\
 & \Xbeta_2  \betavsw{1} = \muswt{2}{\perm_2(1)},  &   &  \Xbeta_2  \betavsw{2} = \muswt{2}{\perm_2(2)} . \nonumber
 \end{align}
If $\perm_2$  is  identical to $\perm_1$, then  the original  values $\betavsw{1}$ and  $\betavsw{2}$ are recovered through:
\begin{eqnarray} \label{betatrue}
\betavsw{1} = \Xbetamat_{1,2} ^{-1}  \left( \begin{array}{c}  \muswt{1}{1}  \\   \muswt{2}{1}   \end{array} \right) =  \left( \begin{array}{r}     2 \\  2   \end{array} \right), \quad
\betavsw{2} =   \Xbetamat _{1,2} ^{-1}   \left( \begin{array}{c}  \muswt{1}{2}  \\   \muswt{2}{2}   \end{array} \right) =  \left( \begin{array}{r}     1 \\  -2   \end{array} \right),
\end{eqnarray}
since  the design matrix
\begin{eqnarray*}
\Xbetamat _{1,2} =\left( \begin{array}{c}   \Xbeta_1  \\   \Xbeta_{2}   \end{array} \right) = \left( \begin{array}{cc}  1  & 0  \\   1 & 1   \end{array} \right)
\end{eqnarray*}
is invertible.
However, as mentioned above,   $\perm_2$  need not be identical to $\perm_1$, in which case   $\perm_2(1)=2, \perm_2(2)=1$, and a second solution
emerges:
\begin{eqnarray*}
\idestar{\betavsw{1}} =  \Xbetamat _{1,2}  ^{-1}   \left( \begin{array}{c}  \muswt{1}{1}  \\   \muswt{2}{2}   \end{array} \right)
=  \left( \begin{array}{r}     2 \\  - 3   \end{array} \right), \quad
\idestar{\betavsw{2}} = \Xbetamat  _{1,2} ^{-1}   \left( \begin{array}{c}  \muswt{1}{2}  \\   \muswt{2}{1}   \end{array} \right)
=  \left( \begin{array}{r}     1 \\   3   \end{array} \right).
\end{eqnarray*}
 Evidently, the group-specific slopes of this second solution are different from the original ones and  the un-identifiability  set $\ideU (\thmod \true)$
  contains two points.
 The two possible solutions are depicted in the left-hand side of Figure~\ref{regmix:data} which also
 shows a balanced sample of $n=100$ observations simulated from   mixture (\ref{mixreg2}) under  \textit{Design~1}.

   For \textit{Design~2},   the first two design points  are as before and 
   a third point is added  with  $\muswt{3}{1}$ and $\muswt{3}{2}$ being identified up to label switching
  according to  a permutation $\perm_3$:
  \begin{align} \label{regperm2}
&\Xbeta_3  \betavsw{1} = \muswt{3}{\perm_3(1)} ,
& &  \Xbeta_3  \betavsw{2} = \muswt{3}{\perm_3(2)} .
\end{align}
As only two different permutations exist for $G=2$,  at least two of the three  permutations $\perm_1$, $\perm_2$, and $\perm_3$  in (\ref{regperm1}) and (\ref{regperm2}) have to be identical (assuming again without loss of generality that  $\perm_1$ is equal to the identity).
Assume, for example, that  $\perm_1= \perm_2$. Then  the true parameters $\betavsw{1}$ and $\betavsw{2}$ are  recovered from $(\muswt{j}{1},\muswt{j}{2}),j=1,2,$   as in  (\ref{betatrue}) and
can be used to uniquely predict $\muswt{3}{1}= \Xbeta_3  \betavsw{1}$ and
$\muswt{3}{2} = \Xbeta_3  \betavsw{2} $ in both groups.
%
Comparing these predictions with  (\ref{regperm2}), it is clear that   $\perm_3(1)=1$ and  $\perm_3(2)=2$, hence
$\perm_1 = \perm_2 = \perm_3$. A similar proof  can be performed for any  pair of identical  permutations $\perm_j = \perm_l, j\neq l$, as long as
the  matrix  $\Xbetamat _{j,l}^\top = (  \Xbeta_j  ^\top \,\,   \Xbeta_{l}^\top  ) $ 
 is invertible and generic identifiability of  \textit{Design~2} follows.

The only possible solution under \textit{Design~2}  is  depicted in the right-hand  side of Figure~\ref{regmix:data} which also shows  a balanced sample of $n=100$ observations simulated from
    mixture (\ref{mixreg2}) under  this design.

 \begin{figure}[t!]
\begin{center}
\begin{tabular}{cc}
\scalebox{0.3}{\includegraphics{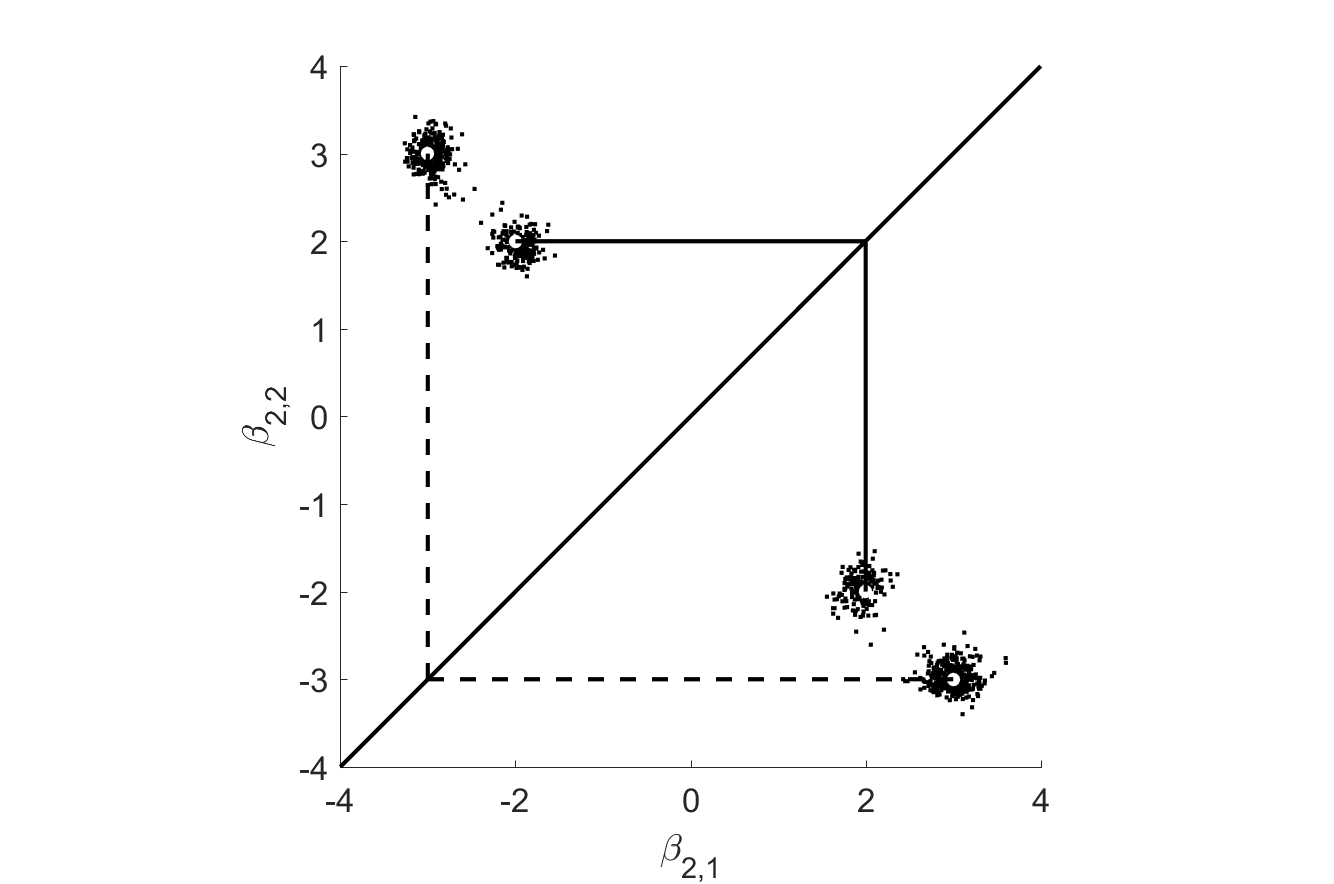}} & \scalebox{0.3}{\includegraphics{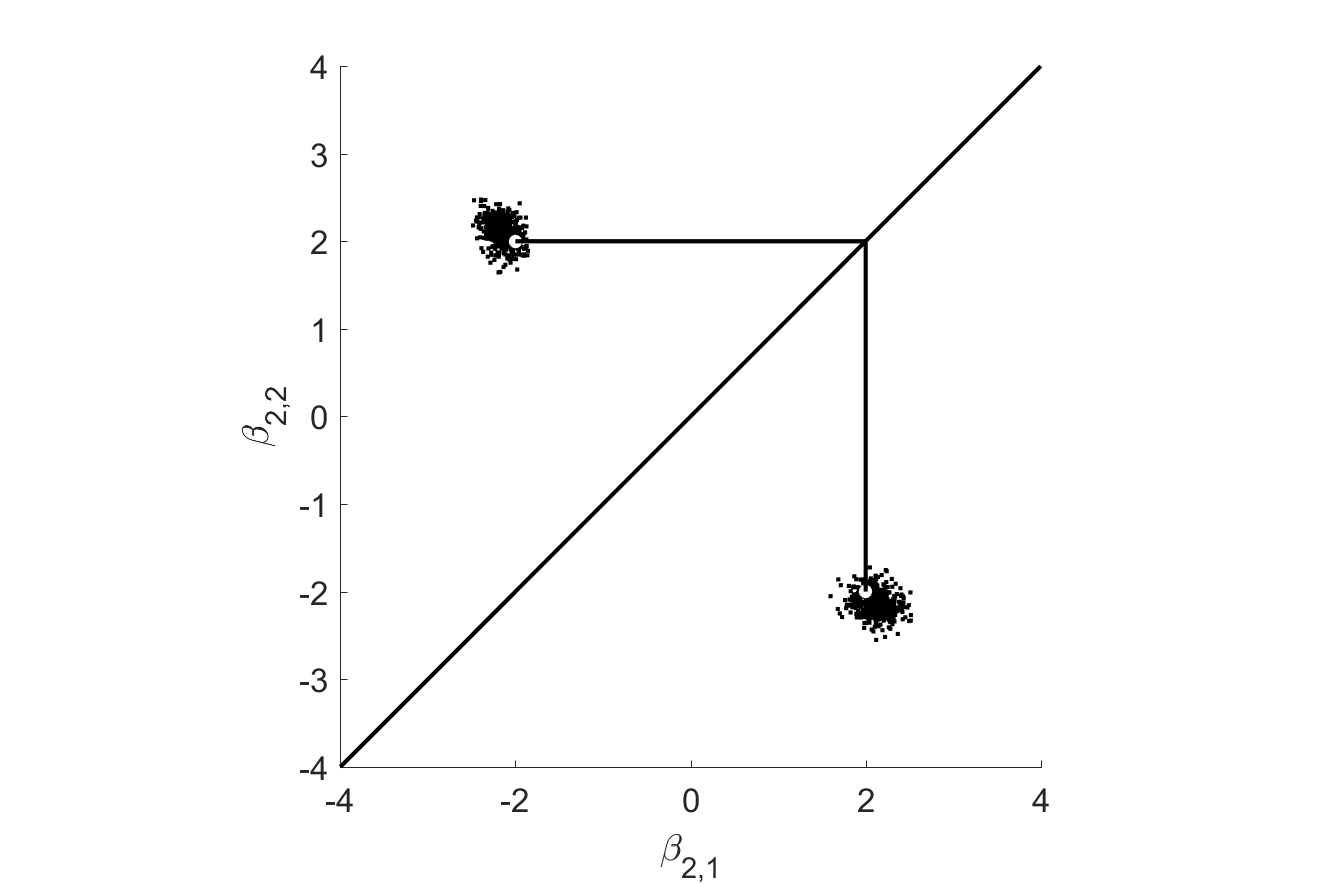}}\\
\scalebox{0.3}{\includegraphics{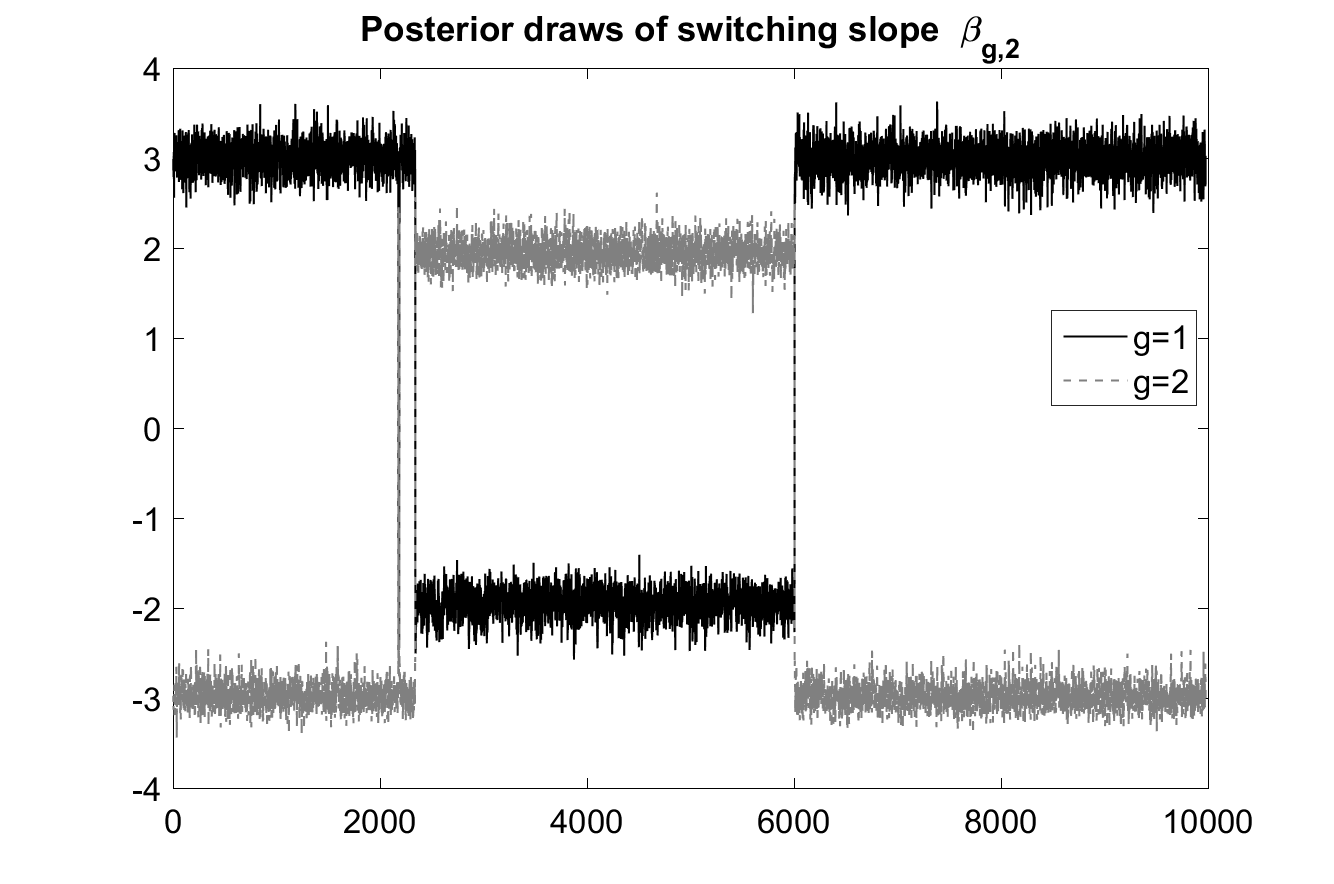}} & \scalebox{0.3}{\includegraphics{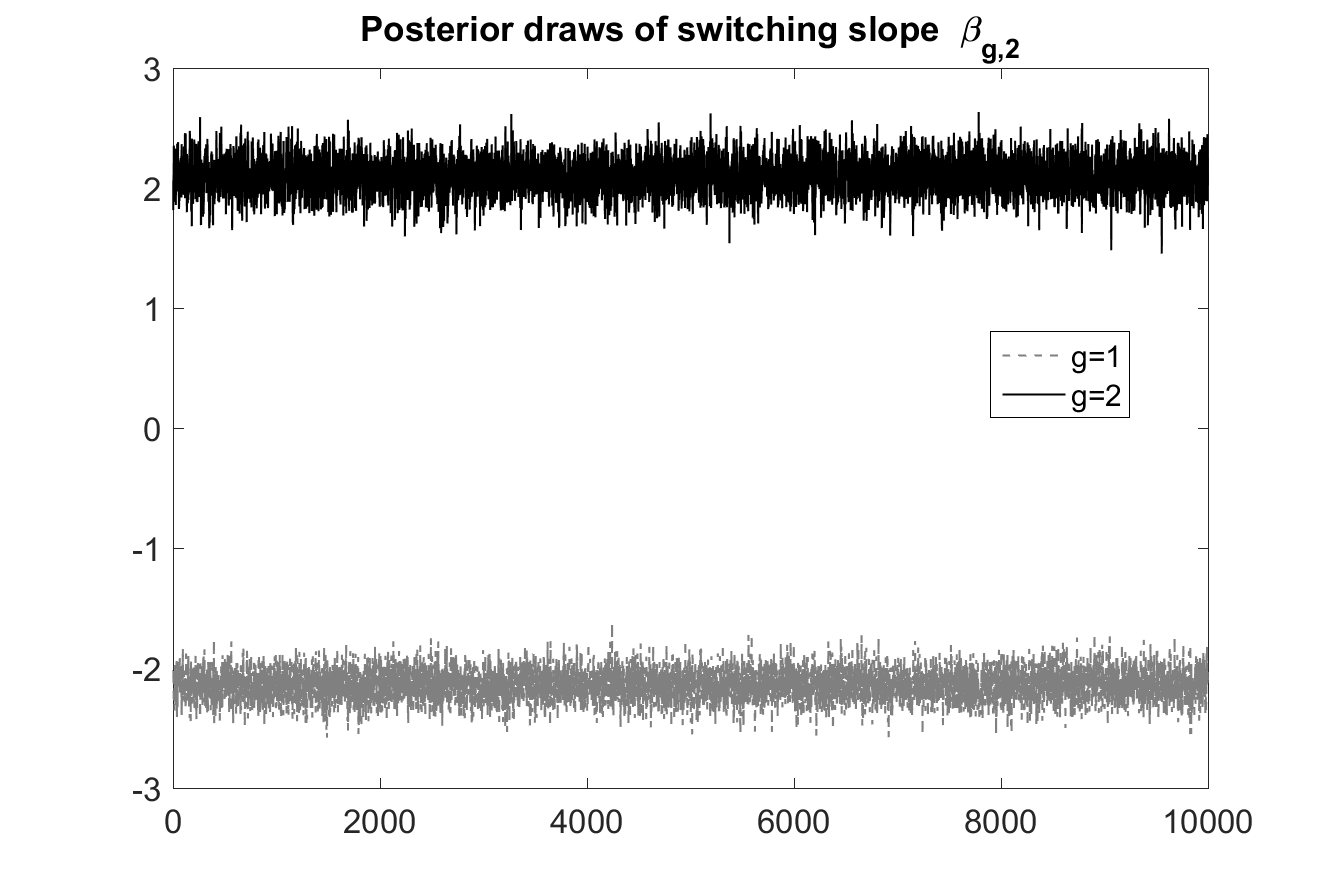}}
\end{tabular}
\end{center}
\vspace{-0.75cm}
\caption{MCMC inference for data simulated from a mixture of two regression models under the generically  un-identified  \textit{Design~1} (left-hand side) and
under  the generically  identified \textit{Design~2} (right-hand side).  Top: scatter plot of  the group-specific slopes $\betacsw{1}{2}$ versus $\betacsw{2}{2}$.  Bottom: posterior draws of  the group-specific slopes  $\betacsw{1}{2}$ and $\betacsw{2}{2}$ after resolving label switching through $k$-means clustering in
the  posterior draws.} 
\label{regmix:mcmc}
\end{figure}

  \paragraph*{MCMC inference for an example: a mixture of regressions model}
  $\:$\\
  $\:$\\
  For further illustration, we perform MCMC inference (based on 10,000 draws after a burn-in of 5,000 iterations)
  for both data sets shown in  Figure~\ref{regmix:data} using random permutation sampling as explained in  Chapter~5, Section~5.2.  We assume that $G=2$ is known,  whereas  all other parameters   in  mixture (\ref{mixreg1}) are unknown.
  Bayesian inference is based on the priors   $(\eta_1,\eta_2) \sim \Dir(4,4)$ and   $\betavsw{g} \sim \Normal(\bfz,100\times \identm),\sigmaerrsw{g} \sim \IG (2.5,1.25 s_y^2 ) $ for $g=1,2$, where $ s_y^2$ is the data variance of $(y_1,\ldots,y_n)$.
 The upper part of  Figure~\ref{regmix:mcmc}
 shows scatter plots  of   the group-specific slopes $\betacsw{1}{2}$ versus $\betacsw{2}{2}$ for both designs which are symmetric  due to label switching.

  As expected from  Chapter~4, Section~4.3,    the posterior draws  shown in the upper  right-hand part   for  the generically identified \textit{Design~2}   concentrate  around  two symmetric modes  corresponding to the true values $(2,-2)$ and $(-2,2) $.
  Label switching is easily resolved by applying $k$-means clustering to the posterior draws,
 see the   lower right-hand part of Figure~\ref{regmix:mcmc}   showing  identified posterior draws of  the group-specific slopes $\betacsw{1}{2}$ and
  $\betacsw{2}{2}$  for this design.

For  \textit{Design~1}, a similar  scatter plot of $\betacsw{1}{2}$ versus $\betacsw{2}{2}$ in the upper  left-hand part   clearly indicates severe identifiability issues.   The posterior draws concentrate  around  four rather than two  modes,  with  two of them being the symmetric modes corresponding to the true values $(2,-2)$  and  $(-2,2)$.  The other  two symmetric modes correspond to the second solution $(-3,3)$ and $(3,-3)$, resulting from generic non-identifiability.
  When we apply $k$-means clustering to these posterior draws to resolve label switching, we obtain the posterior draws of   the group-specific slopes $\betacsw{1}{2}$ and  $\betacsw{2}{2}$
 in the lower left-hand part of Figure~\ref{regmix:mcmc},  showing {\em intra-component label switching} and\index{label switching!intra-component}
    indicating identifiability problems  for this design.
 Since $\betacsw{g}{2}$ switches sign between the two solutions in both groups, it is not possible to recover
  that \lq\lq gender\rq\rq\  has a strong positive  effect on the outcome in one group and a strong  negative effect in the other group.

\subsection{Identifiability for mixtures of experts models}  \label{sec_ideME}

\citet{hen:ide} considers mixtures of regression models where the component sizes can arbitrarily depend on covariates and establishes identifiability results in the case that the joint observations of covariates and dependent variable are assumed to be iid and gives sufficient identifiability conditions for this model.
For such a model,  the covariates are not assumed fixed or to occur for a fixed design, but random with a specific distribution.
 As opposed to this, the ME  model  is defined  conditional on the covariates  without  specific assumptions  concerning their distribution. We will discuss identification for this case.

Consider, as a first example, a simple mixture of experts model of  $G$ univariate Gaussian distributions for $i=1, \ldots, n$
outcomes $y_i$ arising from $G$  different groups,
\begin{eqnarray} \label{mixexpreg1}
y_i  | \Xbetatilde_i \sim \sum_{g=1}^G \eta_g ( \Xbetatilde_i) \phi (y|  \mu_{g},\sigma^2_{g})
\end{eqnarray}
where the group weights  $\eta_g$ depend  on a covariate  $ \Xbetatilde_i$ with   group-specific  regression parameters, i.e.:
\begin{eqnarray}  \label{MNL}
\log \left[ \frac{\eta_g  ( \Xbetatilde_i)}{\eta_{g_0}  ( \Xbetatilde_i)}  \right] = \Xbetatilde_i  \gamma_{g},  \quad g=1,\ldots, G,
\end{eqnarray}
with baseline $g_0$, where  $\gamma_{g_0}=\bfz$. Assume that the component densities differ, i.e. $\theta_g \neq  \theta_g'$, for $g \neq g'$, where   $\theta_g=(\mu_g, \sigma^2_{g})$.

For each fixed design point $\Xbeta=\Xbetatilde_i$, (\ref{mixexpreg1}) is a standard finite Gaussian mixture and therefore generically identified.
Therefore, if the identity
\begin{eqnarray}
\sum_{g=1}^G \eta_g ( \Xbeta) \phi (y| \mu_{g},\sigma^2_{g})  =
\sum_{g=1}^G \idestar{\eta}_g ( \Xbeta) \phi (y|  \idestar{\mu}_{g},\idestararg{\sigma_g}{2}) , \label{eqswreg1}
\end{eqnarray}
holds, then  the two mixtures  are related to each other by relabelling, i.e.\index{relabelling}
 $\idestar{\mu}_{g}=\mu_{\perm_x(g)} ,
\idestararg{\sigma_g}{2}=\sigma^2_{\perm_x(g)}$,  and $ \idestar{\eta}_g ( \Xbeta ) =\eta_{\perm_x(g)} (  \Xbeta ) $ for $g=1,\ldots, G$,  for some
permutation $\perm_x \in \mathfrak{S}(G)$.
As opposed to mixtures of regression models, one can show that  $\perm_x \equiv \perm_\star$ for all covariate  values  $\Xbeta$.

Assume that $\perm_{x_i}\neq  \perm_{x_j} $ for two covariates  $\Xbetatilde_i\neq  \Xbetatilde_j$ and assume,  without loss
of generality, that $\perm_{x_i}$ is equal to the identity. Consider first the case of $G=2$. Then (\ref{eqswreg1}) implies for $\Xbeta=\Xbetatilde_i $:
\begin{eqnarray*}
 \idestar{ \theta}_1  =  \left( \begin{array}{l}  \idestar{\mu}_{1} \\   \idestararg{\sigma_1}{2}   \end{array} \right) =   \theta_1,  \quad
\idestar{ \theta}_2 =  \left( \begin{array}{l}  \idestar{\mu}_{2} \\   \idestararg{\sigma_2}{2}   \end{array} \right) =   \theta_2,
\end{eqnarray*}
whereas  for $\Xbeta=\Xbetatilde_j $:
\begin{eqnarray*}
\idestar{ \theta}_1  =  \left( \begin{array}{l}  \idestar{\mu}_{2} \\   \idestararg{\sigma_2}{2}   \end{array} \right) =   \theta_2,  \quad
  \idestar{\theta}_2  =  \left( \begin{array}{l}  \idestar{\mu}_{1} \\   \idestararg{\sigma_1}{2}   \end{array} \right) =   \theta_1,
\end{eqnarray*}
contradicting the assumptions that  $ \theta_1 \neq  \theta_2$.  A similar proof is possible for $G>2$, where  the assumption
$\perm_{x_i}\neq  \perm_{x_j} $ (assuming again that  $\perm_1$ is equal to the identity) implies
 for    $\Xbeta=\Xbetatilde_i $ that   $\idestar{\theta}_{g}  =  \theta_{g}$ for all components  $g=1, \ldots,G$,   whereas  for $\Xbeta=\Xbetatilde_j $
  at least one component $g_i$ exists with  $\idestar{\theta}_{g_i} =  \theta_{g_j}$, where  $g_j= \perm_{x_j}(g_i) \neq g_i$.  Hence,
$ \theta_{g_i}= \theta_{g_j}$,  which contradicts the assumptions that  $ \theta_{g_i} \neq  \theta_{g_j}$.
This implies that  $\perm_x \equiv \perm_\star$ for all covariate  values  $\Xbeta$.

Therefore,  the  weight distribution $\eta_{1} (x), \ldots, \eta_{G} (x) $
 is identified up to relabelling the components and
  identification depends on whether  $\gamma_{g}$ can be recovered from  the corresponding MNL model (\ref{MNL}) given  the design matrix $\Xbetamat$,  constructed row-wise  from the covariates $\Xbetatilde_i, i=1,\ldots, n$.  Standard conditions for identification  in a MNL model apply, e.g.  that  $(\Xbetamat^\top\Xbetamat) ^{-1}$ exists \citep{mcc-nel:gen}. It is well-known that   identification in a logit  and more generally in a MNL  model  fails  under  complete separation, see e.g.~\citet{hei:com}.
  Hence, a situation where a  mixture of experts model is not generically identified  occurs, if
  certain clusters do not share covariate  values with other clusters, see Example~4.2  in  \citet{hen:ide} for illustration.
  A rather strong condition ensuring generic identifiability for this type of models is an extended coverage condition \citep{hen:ide} requiring that the number of clusters $G$ is exceeded by the minimum number of distinct  $q$-dimensional hyperplanes  needed to cover the covariate values (excluding the constant) for {\em each} cluster.


Similar arguments as above apply in general for simple mixtures of experts models of  $G$ probability distributions,
\begin{eqnarray} \label{mixexgen2}
y_i  \sim \sum_{g=1}^G \eta_g ( \Xbetatilde_i) p(y_i|\theta_g).
\end{eqnarray}
Provided that the parameters in the MNL model (\ref{MNL}) are identified,
it can be shown that a  mixture of experts model is generically identified, if the corresponding
standard finite mixture distribution is generically identified. In this  case,   any other mixture representation (\ref{mixexgen2}) with parameters
$\idestar{\theta}_g$ and  $\idestar{\eta}_g (\Xbetatilde_i)$
 is identified up to (the same) label switching according\index{label switching} to a permutation $\perm$  for all possible  values  $\Xbetatilde_i$: $\idestar{\theta}_g= \theta_{\perm(g)}$ and  $\idestar{\eta}_g (\Xbetatilde_i) =\eta_{\perm(g)} ( \Xbetatilde_i ) $  for $g=1,\ldots, G$.

It follows that  mixtures of experts of multivariate
Gaussian distributions (as considered in  Section~\ref{sec:illustrativeeg}) and   Poisson distributions, among many others, are generically identified,
provided that parameters in the MNL model (\ref{MNL}) are identified.
Since a standard finite mixture model is that special case of a  mixture of experts model where  $\Xbetatilde_i \equiv 1$ is equal to the intercept,
special care must be exercised when  the underlying standard finite mixture distribution is  generically unidentified, as might be the case
when modelling discrete data.  It is interesting to note that  including $\Xbetatilde_i$ into the weight function  $\eta_g ( \Xbetatilde_i)$ in mixtures of experts models is possible for models where including $\Xbetatilde_i$ in the component density $p(y_i|\Xbetatilde_i,\theta_g)$ yields a generically non-identified model, an example being the regressor $\Xbetatilde_i =( 1 \,\, d_i)$, where $d_i$ is a 0/1 dummy variable, see Section~\ref{sec13:ide}.

  The situation gets rather complex, when covariates $\Xbetatilde_i$  (or subsets of these) are included as regressors both in
 the outcome  distribution $p(y_i |\Xbetatilde_i, \theta_g)$ as well as in the weight distribution $\eta_g ( \Xbetatilde_i)$.
The presence of  a covariate  $\Xbetatilde_i $ in $ \eta_g ( \Xbetatilde_i)$ could introduce high discriminative power among the groups  and
might lead to identification of mixture of regression models  which are not identified, if  $\eta_g$ is assumed to be independent of the covariates.
 To our knowledge, generic identification for  general mixtures of experts models  has not been studied
 systematically and would be an interesting venue for future research.

 As it is, the only way to investigate, if the chosen mixture  model
 suffers from identifiability problems is to analyze the results obtained from fitting these models to the data carefully. As the examples
 in Section~\ref{sec13:ibin} and ~\ref{sec13:ide} have
 shown, weird behaviour of  the MCMC draws in a Bayesian framework   are often a sign of identifiability problems.
 On the other hand, marginal posterior  concentration around pronounced modes, verified for instance through appropriate
 scatter plots of MCMC draws for the parameters of interest, indicates that identification  might not be an issue for  that specific application.

\section{Concluding Remarks}

This chapter has outlined the definition, estimation and application of ME models in a number of settings clearly demonstrating their utility as an analytical tool. Their demonstrated use to cluster observations, and to appropriately capture heterogeneity in cross sectional data, provides only a glimpse of their potential flexibility and utility in a wide range of settings. The ability of ME models to jointly model response and concomitant variables provides deeper and more principled insight into the relations between such data in a mixture model based analysis.

On a cautionary note however, when an ME model is employed as an analytic tool, care must be exercised in how and where covariates enter the ME model framework. The interpretation of the analysis fundamentally depends on which of the suite of ME models is invoked.
Further, as outlined herein, the identifiability of an ME model must be carefully considered; establishing identifiability for ME models is an outstanding, challenging problem.\index{mixtures!of experts|)}

 \bibliographystyle{authordate1}

\bibliography{GormleyFruhwirthSchnatter}
\end{document}

%% file: tikzlibrarybayesnetcode.tex
\usetikzlibrary{shapes}
\usetikzlibrary{fit}
\usetikzlibrary{chains}
\usetikzlibrary{arrows}

\tikzstyle{latent} = [circle,fill=white,draw=black,inner sep=1pt,
minimum size=20pt, font=\fontsize{10}{10}\selectfont, node distance=1]
\tikzstyle{obs} = [latent,fill=gray!25]
\tikzstyle{const} = [rectangle, inner sep=0pt, node distance=1]
\tikzstyle{factor} = [rectangle, fill=black,minimum size=5pt, inner
sep=0pt, node distance=0.4]
\tikzstyle{det} = [latent, diamond]

\tikzstyle{plate} = [draw, rectangle, rounded corners, fit=#1]
\tikzstyle{wrap} = [inner sep=0pt, fit=#1]
\tikzstyle{gate} = [draw, rectangle, dashed, fit=#1]

\tikzstyle{caption} = [font=\footnotesize, node distance=0] %
\tikzstyle{plate caption} = [caption, node distance=0, inner sep=0pt,
below left=0pt and 0pt of #1.south east] %
\tikzstyle{factor caption} = [caption] %
\tikzstyle{every label} += [caption] %

\tikzset{>={triangle 45}}

\newcommand{\edge}[3][]{ %
  \foreach \x in {#2} { %
    \foreach \y in {#3} { %
      \draw[->,#1] (\x) -- (\y) ;%
    } ;
  } ;
}

\newcommand{\plate}[4][]{ %
  \node[wrap=#3] (#2-wrap) {}; %
  \node[plate caption=#2-wrap] (#2-caption) {#4}; %
  \node[plate=(#2-wrap)(#2-caption), #1] (#2) {}; %
}

